\documentclass[12pt,preprint]{aastex}

\newcommand{\teff}{T_\mathrm{eff}}

\bibpunct{(}{)}{,}{a}{}{,}

\usepackage{array}
\usepackage{tabularx}
\usepackage{natbib}
\shorttitle{STELLAR AGES AND CONVECTIVE CORES IN {\it KEPLER} TARGETS}
\shortauthors{SILVA AGUIRRE ET AL.}

\begin{document}

\title{Stellar ages and convective cores in field main-sequence stars: first asteroseismic application to two \textit{Kepler} targets}
\author{V. Silva Aguirre\altaffilmark{1,2,3}, S. Basu\altaffilmark{4}, I.~M. Brand\~{a}o\altaffilmark{5}, J. Christensen-Dalsgaard\altaffilmark{1,3}, S. Deheuvels\altaffilmark{4,6,7}, G. Do\u{g}an\altaffilmark{8}, T.~S. Metcalfe\altaffilmark{9}, A.~M. Serenelli\altaffilmark{10,3}, J. Ballot\altaffilmark{6,7}, W.~J. Chaplin\altaffilmark{11,1,3}, M.~S. Cunha\altaffilmark{5}, A. Weiss\altaffilmark{2}, T. Appourchaux\altaffilmark{12}, L. Casagrande\altaffilmark{13}, S. Cassisi\altaffilmark{14}, O.~L. Creevey\altaffilmark{15}, R.~A. Garc\'\i a\altaffilmark{16,3}, Y. Lebreton\altaffilmark{17,18}, A. Noels\altaffilmark{19}, S.~G. Sousa\altaffilmark{5}, D. Stello\altaffilmark{20}, T.~R. White\altaffilmark{20}, S.~D Kawaler\altaffilmark{21}, and H. Kjeldsen\altaffilmark{1}}
\altaffiltext{1}{Stellar Astrophysics Centre, Department of Physics and Astronomy, Aarhus University, Ny Munkegade 120, DK-8000 Aarhus C, Denmark}
\altaffiltext{2}{Max Planck Institute for Astrophysics, Karl-Schwarzschild-Str. 1, 85748, Garching bei M\"{u}nchen, Germany}
\altaffiltext{3}{Kavli Institute for Theoretical Physics, Santa Barbara, California 93106, USA}
\altaffiltext{4}{Department of Astronomy, Yale University, P.O. Box 208101, New Haven, CT 06520-8101, USA}
\altaffiltext{5}{Centro de Astrof\'{\i}sica and Faculdade de Ci\^encias, Universidade do Porto, Rua das Estrelas, 4150-762 Porto, Portugal}
\altaffiltext{6}{Institut de Recherche en Astrophysique et Plan\'etologie, CNRS, 14 avenue Edouard Belin, 31400 Toulouse, France}
\altaffiltext{7}{Universit\'e de Toulouse, UPS-OMP, IRAP, Toulouse, France}
\altaffiltext{8}{High Altitude Observatory, NCAR, P.O. Box 3000, Boulder, CO 80307, USA}
\altaffiltext{9}{Space Science Institute, Boulder, CO 80301, USA}
\altaffiltext{10}{Instituto de Ciencias del Espacio (CSIC-IEEC), Facultad de  Ci\`encies, Campus UAB, 08193 Bellaterra, Spain}
\altaffiltext{11}{School of Physics and Astronomy, University of Birmingham, Birmingham, B15 2TT, UK}
\altaffiltext{12}{Institut d'Astrophysique Spatiale Universit\'e Paris Sud - CNRS (UMR8617) Batiment 121, F-91405 ORSAY Cedex}
\altaffiltext{13}{Research School of Astronomy and Astrophysics, Mount Stromlo Observatory, The Australian National University, ACT 2611, Australia}
\altaffiltext{14}{INAF - Astronomical Observatory of Teramo, Via M. Maggini, sn, 64100 Teramo, Italy}
\altaffiltext{15}{Universit\'e de Nice Sophia-Antipolis, Laboratoire Lagrange, CNRS, Nice, 06300, France.}
\altaffiltext{16}{Laboratoire AIM, CEA/DSM-CNRS-Universit\'e Paris Diderot; IRFU/SAp, Centre de Saclay, 91191 Gif-sur-Yvette Cedex, France}
\altaffiltext{17}{Observatoire de Paris, GEPI, CNRS UMR 8111, 92195, Meudon, France}
\altaffiltext{18}{Institut de Physique de Rennes, UniversitŽ de Rennes 1, CNRS UMR 6251, 35042, Rennes, France}
\altaffiltext{19}{Institute of Astrophysics and Geophysics, University of Li\`ege, Belgium}
\altaffiltext{20}{Sydney Institute for Astronomy (SIfA), School of Physics, University of Sydney, NSW 2006, Australia}
\altaffiltext{21}{Department of Physics and Astronomy, Iowa State University, Ames, IA, USA 50014}
\begin{abstract}
Using asteroseismic data and stellar evolution models we make the first detection of a convective core in a \textit{Kepler} field main-sequence star, putting a stringent constraint on the total size of the mixed zone and showing that extra mixing beyond the formal convective boundary exists. In a slightly less massive target the presence of a convective core cannot be conclusively discarded, and thus its remaining main-sequence life time is uncertain. Our results reveal that best-fit models found solely by matching individual frequencies of oscillations corrected for surface effects do not always properly reproduce frequency combinations. Moreover, slightly different criteria to define what the best-fit model is can lead to solutions with similar global properties but very different interior structures. We argue that the use of frequency ratios is a more reliable way to obtain accurate stellar parameters, and show that our analysis in field main-sequence stars can yield an overall precision of~1.5\%,~4\%,~and~10\% in radius, mass and age, respectively. We compare our results with those obtained from global oscillation properties, and discuss the possible sources of uncertainties in asteroseismic stellar modeling where further studies are still needed.
\end{abstract}

\section{Introduction}\label{s:intro}
A detailed comprehension of the physical processes taking place in deep stellar interiors is of paramount importance for an accurate description of stellar populations \citep[e.g.,][]{Chiosi:1992gp}. Assumptions and parameterizations used to represent processes like convection, rotation, overshooting and microscopic diffusion directly impact the quantities normally used to characterize different types of stars, such as effective temperatures, colors, surface gravities and composition \citep[see e.g.,][and references therein]{1978A&A....65..281R,1988PASP..100..314V,1993ASPC...40..246M,Maeder:2000br}. As a consequence it is extremely difficult to determine reliable stellar ages of field main-sequence stars, where uncertainties are well above the 20\% level \citep{Soderblom:2010kr}.

One aspect of critical relevance for dating main sequence stars is the existence and size of a convective core. It has been long known \citep[e.g.,][]{Maeder:1974wc} that reproducing the observed hook-like feature in the Color-Magnitude Diagram (CMD) of intermediate-age clusters requires that stars at the turn-off phase possessed a convective core during their main-sequence lifetime. Since the metallicity of cluster stars is assumed to be fixed and homogeneous, the age of the cluster can be estimated by determining the best fit of isochrones to the observed CMD \citep[e.g.,][]{1985ApJS...58..561V,1997ApJ...484..986S}. The great caveat of this technique is that the critical mass at which a convective core appears in evolutionary models, normally around 1.1~$M_\sun$, strongly depends on the input physics \citep[see e.g.,][]{ChristensenDalsgaard:2010er,Magic:2010iz}. Moreover, the total size of the mixed central region in stellar models can be controlled by a parameter of mixing efficiency beyond the formal convective boundaries of the core; this is calibrated to match observations of clusters and is again dependent on the rest of the input physics used to construct isochrones \citep[e.g.,][]{Maeder:1991to,Pietrinferni:2004im}. Therefore, we currently lack a reliable estimation of the size of convective cores in main-sequence stars.

An exciting approach to overcome these impediments comes from the possibility of piercing the surface of stars by studying their pulsations, and the exquisite data currently being obtained by the \textit{Kepler} mission provide an excellent opportunity to achieve this goal \citep{ChristensenDalsgaard:2007ku,Borucki:2009bb}. Several theoretical studies of how asteroseismology can yield information about stellar cores have been carried out \citep[e.g.,][]{Audard:1994wr,Mazumdar:2001kc,Mazumdar:2006fn,Cunha:2007gs,Cunha:2011dh,SilvaAguirre:2011jz}, but the paucity of data for main-sequence targets prior to the launch of \textit{Kepler} allowed very few investigations of this type in stars other than the Sun. In fact, all previous data-based asteroseismic studies of stellar cores properties \cite[such as][]{Miglio:2005is,Deheuvels:2010fn,DeMelauner:2010bp} relied on ground-based observations of much shorter time coverage and lower quality than the multi-month \textit{Kepler} observations.

From the hundreds of main-sequence targets for which the \textit{Kepler} mission has provided asteroseismic data \citep{Chaplin:2011bi}, individual frequencies have now been determined for dozens of them from time-series spanning several months \citep[e.g.,][]{Appourchaux:2012kd}. We carry out here the first investigation using {\it Kepler} data on two of these targets with the aim of detecting convective cores, constraining their size, determining accurate stellar parameters, and narrowing down the age uncertainties compared to those obtained by fitting stellar tracks.

To accomplish this, we proceed as follows: in Section~\ref{s:targs} we describe the available seismic and spectroscopic observations, which are used in Section~\ref{s:dir_and_grid} to obtain initial stellar parameters from global asteroseismic fitting techniques. The large uncertainties obtained from this analysis, particularly in age, lead us to consider diagnostic tools sensitive to the deep interior of stars, presented in Section~\ref{s:ratios}. Several teams performed detailed modeling of the targets using different evolution and pulsation codes, described in Section~\ref{s:modeling}. The results of this detailed analysis are given in Section~\ref{s:results}, together with the derived stellar parameters. We discuss our findings, as well as other possible sources of uncertainties, and give closing remarks in Section~\ref{s:disc}.
%
\section{Target selection and observations}\label{s:targs}
\begin{table*}[!ht]\scriptsize
\centering
\caption{Parameters determined from the observations of the \textit{Kepler} mission, ground based photometry, and spectroscopy for both stars. Luminosity of Perky determined from parallax measurements and bolometric corrections. Kepler magnitude K$_\mathrm{p}$ as determined by \citet{Brown:2011dr}.}
\label{tab:targ}
\begin{tabular}{c c c c c c c c c}
\hline\hline
\noalign{\smallskip}
Star & KIC & K$_\mathrm{p}$ & $\nu_{\mathrm{max}} \, (\mu$Hz) & $\langle\Delta\nu\rangle \, (\mu$Hz)  & $T_\mathrm{eff}$ (K) & $\log\,g$& [Fe/H] & $\log\,(L/L_\odot)$\\
\noalign{\smallskip}
\hline
\noalign{\smallskip}
Perky & 6106415 & 7.18 &$2210\pm50$ & $104\pm0.5$ & $6000\pm200$ & $4.27\pm0.1$ & $-0.09\pm0.1$ & $0.26\pm0.04$ \\
\noalign{\smallskip}
Dushera & 12009504 & 9.32 & $1833\pm40$ & $88\pm0.6$ & $6200\pm200$ & $4.30\pm0.2$ & $0.0\pm0.15$ & ---\\
\noalign{\smallskip}
\hline
\end{tabular}
\end{table*}
Based on the full ensemble of stars with 9 months of \textit{Kepler} observations available, we initially selected targets with seismic characteristics belonging to the main-sequence phase \citep[no signature of mixed-modes, see][for a thorough explanation]{Bedding:2011wl}. Since we are interested in detecting convective cores, we further discarded the targets with global seismic properties suggesting a mass value lower than solar, and finally chose two stars from the remaining targets with a large number of reliable individual frequencies determined \citep[see][]{Appourchaux:2012kd}. These are dubbed Perky (KIC~6106415) and Dushera (KIC~12009504).

The power spectrum of solar-like oscillators is modulated by a Gaussian-like envelope, and presents a near-regular pattern of overtone frequencies \citep[e.g.,][]{Chaplin:2011bi}. Two parameters can be readily extracted from it, namely the frequency of maximum power of oscillation $\nu_{\mathrm{max}}$ and $\langle\Delta\nu\rangle$, the average separation between the consecutive p-mode overtones:
\begin{equation}\label{eq:large}
\Delta\nu_{\ell}(n)  =  \nu_{n,\ell}-\nu_{n-1,\ell}\,,
\end{equation}
where $\nu_{n,\ell}$ is the mode frequency of angular degree $\ell$ and radial order $n$. These are usually referred to as the {\it global} oscillation parameters.

Values of $\langle\Delta\nu\rangle$ and $\nu_{\mathrm{max}}$ were obtained for both targets as described by \citet{Verner:2011gw} using time series prepared from short-cadence data \citep{Gilliland:2010iq,Garcia:2011ho}. Individual frequencies were extracted in the manner explained by \citet{Appourchaux:2012kd}, resulting in 33 oscillation modes for Perky and 34 for Dushera. These are listed in Table~\ref{tab:frq_perk} and in Table~A.53 of \citet{Appourchaux:2012kd} for Perky and Dushera, respectively.

Effective temperature $\teff$, surface gravity $\log\,g$ and iron abundance [Fe/H] were available for both targets from high-resolution spectroscopic observations analyzed with the VWA software \citep[][and references therein]{Bruntt:2012hz}. The case of Dushera presented a challenge: a set of low-resolution spectra was analyzed with a standard method based on the equivalent widths of iron lines using ARES and MOOG \citep[see e.g.,][]{Sousa:2007gn}. This method, described in detail in \citet{Sousa:2008do}, yielded atmospheric parameter values that were not consistent with those obtained from VWA. Although one might be tempted to disregard the results from the lower quality data, estimations of $\teff$ for Dushera using the InfraRed Flux Method \citep[IRFM,][]{Casagrande:2010hj,SilvaAguirre:2012du} suggest a value that is consistent with that obtained by ARES$+$MOOG. Since a thorough comparison of these spectroscopic methods is beyond the scope of this paper, we adopted average values and large uncertainties in the observed parameters that encompassed both determinations.

For the case of Perky only the results from \citet{Bruntt:2012hz} were available and the IRFM $\teff$ determination is consistent with this value. However, based on our experience with Dushera, we decided to use more conservative uncertainties than those quoted by \citet{Bruntt:2012hz}. Since parallax measurements are also available for this star, we estimated its luminosity using the {\it Hipparcos} data \citep{VanLeeuwen:2007dc} and bolometric corrections \citep{1996ApJ...469..355F,Torres:2010gd}. The complete set of observational constraints considered in this study is given in Table~\ref{tab:targ}.

An initial determination of ages can be made using these spectroscopic parameters and stellar isochrones following a bayesian analysis \citep{Serenelli:2013fz}. For the values quoted in Table~\ref{tab:targ}, this technique yields age uncertainties of 50\% and 60\% for Perky and Dushera, respectively. Even when reducing the errors in $\teff$ to 100~K and considering and error in $\log\,g$ of 0.1~dex for both targets, the obtained uncertainties only decrease to 36\% and 50\%. Ages determined with a precision of $\sim$20\% can only be achieved with this technique for targets with accurately determined observational constraints, which are not available for these relatively faint field main-sequence stars (see Section~5.2 in \citet{Serenelli:2013fz}). In fact, age results for main-sequence intermediate-mass field stars have often uncertainties of 20-50\% when determined from a single stellar evolution database, that is, before systematics are taken into account (see Section~4.1 in \citet{Soderblom:2010kr}).
\section{Global asteroseismic analysis}\label{s:dir_and_grid}
\begin{table*}[!ht]\tiny
\centering
\caption{Stellar parameters as determined from each pipeline for the two targets. Also shown are the values obtained with the direct method. See text for details.}
\begin{tabular}{c c c c c c c c c c c c}
\hline
\noalign{\smallskip}
Star & Parameter & YY & Dotter & Marigo & YREC & RADIUS & GARSTEC & SEEK & RadEx10 & BaSTI & Direct\\
\noalign{\smallskip}
\hline\hline
\noalign{\smallskip}
Perky & $M(M_\odot$)& $1.06^{+0.09}_{-0.07}$ & $1.06^{+0.08}_{-0.07}$ & $1.12^{+0.05}_{-0.06}$ & $1.09^{+0.07}_{-0.06}$ & $1.10\pm0.08$ & $1.11^{+0.10}_{-0.10}$ & $1.03^{+0.10}_{-0.10}$ & $1.08\pm0.09$ &$1.07^{+0.06}_{-0.10}$ & $1.10\pm0.07$ \\
\noalign{\smallskip}
         & $R(R_\odot$)& $1.21^{+0.03}_{-0.03}$ & $1.21^{+0.03}_{-0.03}$ & $1.23^{+0.02}_{-0.02}$ & $1.22^{+0.03}_{-0.02}$ & $1.22\pm0.03$ & $1.24^{+0.01}_{-0.04}$ & $1.20^{+0.04}_{-0.03}$ & $1.22\pm0.03$ &  $1.22^{+0.02}_{-0.04}$ & $1.23\pm0.02$ \\
\noalign{\smallskip}
              & $\log\,g$ & $4.30^{+0.01}_{-0.01}$ & $4.30^{+0.01}_{-0.01}$ & $4.31^{+0.01}_{-0.02}$ & $4.31^{+0.01}_{-0.01}$ & $4.30\pm0.01$ & $4.30^{+0.01}_{-0.01}$& $4.29^{+0.02}_{-0.02}$ & $4.30\pm0.01$ &  $4.30^{+0.01}_{-0.01}$ & $4.30\pm0.01$ \\
\noalign{\smallskip}
         & Age (Gyr)& $5.73^{+2.38}_{-2.44}$ & $6.00^{+2.44}_{-2.18}$ & $3.71^{+1.80}_{-1.22}$ & $6.21^{+2.24}_{-2.15}$ & $6.77\pm2.20$ & $4.21^{+3.36}_{-2.21}$ & $5.52^{+4.54}_{-3.66}$ & $6.37\pm3.18$ & $5.80^{+3.70}_{-2.20}$ &$--$ \\
\noalign{\smallskip}
\hline
\noalign{\smallskip}
Dushera & $M(M_\odot$) & $1.25^{+0.08}_{-0.10}$ & $1.21^{+0.09}_{-0.08}$ & $1.23^{+0.07}_{-0.05}$ & $1.27^{+0.10}_{-0.10}$ & $1.21\pm0.10$ & $1.26^{+0.09}_{-0.09}$ & $1.16^{+0.14}_{-0.13}$ & $1.26\pm0.09$ & $1.19^{+0.06}_{-0.08}$ & $1.22\pm0.08$ \\
\noalign{\smallskip}
              & $R(R_\odot$) & $1.43^{+0.04}_{-0.04}$ & $1.42^{+0.03}_{-0.03}$ & $1.42^{+0.03}_{-0.02}$ & $1.44^{+0.04}_{-0.04}$ & $1.41\pm0.05$ & $1.44^{+0.03}_{-0.03}$ & $1.41^{+0.05}_{-0.05}$ & $1.44\pm0.04$ & $1.41^{+0.02}_{-0.03}$ & $1.42\pm0.03$ \\
\noalign{\smallskip}
              & $\log\,g$ & $4.22^{+0.01}_{-0.01}$ & $4.22^{+0.01}_{-0.01}$ & $4.22^{+0.01}_{-0.01}$ & $4.22^{+0.01}_{-0.01}$ & $4.22\pm0.01$ & $4.22^{+0.01}_{-0.01}$ & $4.21^{+0.02}_{-0.02}$ & $4.22\pm0.01$ &  $4.22^{+0.01}_{-0.01}$ & $4.22\pm0.01$ \\
\noalign{\smallskip}
              & Age (Gyr) & $3.21^{+1.83}_{-1.20}$ & $3.82^{+1.68}_{-1.36}$ & $2.94^{+1.06}_{-1.05}$ & $3.78^{+2.13}_{-1.54}$ & $6.43\pm2.06$ & $2.89^{+1.61}_{-1.07}$ & $3.75^{+3.83}_{-2.38}$ & $3.74\pm1.71$ & $4.05^{+1.81}_{-1.05}$ & $--$ \\
\noalign{\smallskip}
\hline
\end{tabular}
\label{tab:grid}
\end{table*}
Using the parameters given in Table~\ref{tab:targ}, it is possible to apply global asteroseismic techniques and determine the stellar properties of the targets. $\langle\Delta\nu\rangle$ approximately scales as the square root of the mean density \citep[e.g.,][]{Ulrich:1986ge}, while $\nu_\mathrm{max}$ seems to be related to the acoustic cutoff frequency of the atmosphere and thus to the surface gravity and effective temperature of the star \citep[e.g.,][]{Brown:1991cv,Kjeldsen:1995tr,Belkacem:2011hm}. Based on these dependencies, two scaling relations anchored at the solar parameters can be written
\begin{equation}\label{eqn:mass} 
\frac{M}{M_\odot} \simeq \left(\frac{\nu_{\mathrm{max}}}{\nu_{\mathrm{max},\odot}}\right)^{3} \left(\frac{\langle\Delta\nu\rangle}{\Delta\nu_\odot}\right)^{-4}\left(\frac{T_\mathrm{eff}}{T_{\mathrm{eff},\odot}}\right)^{3/2}, 
\end{equation}
\begin{equation}\label{eqn:rad} 
\frac{R}{R_\odot} \simeq \left(\frac{\nu_{\mathrm{max}}}{\nu_{\mathrm{max},\odot}}\right) \left(\frac{\langle\Delta\nu\rangle}{\Delta\nu_\odot}\right)^{-2}\left(\frac{T_\mathrm{eff}}{T_{\mathrm{eff},\odot}}\right)^{1/2}. 
\end{equation}
Here, $T_{\mathrm{eff},\odot}= 5777\,\rm K$, $\Delta\nu_\odot = 135.1\pm0.1\,\rm \mu Hz$ and $\nu_{\mathrm{max},\odot}= 3090\pm30\,\rm \mu Hz$ are the observed values in the Sun \citep{Huber:2011be}. Provided a measurement of $\teff$ is available, these relations give a determination of mass and radius independent of stellar models \citep[the {\it direct method}, see e.g.,][]{Miglio:2009hz,Chaplin:2011bi,SilvaAguirre:2011es}. It is also possible to search for a best fit to these parameters within pre-calculated sets of evolutionary tracks, where chemical composition can be taken into account \citep[the {\it grid-based method}, e.g.,][]{Stello:2009fg,Quirion:2010is,Gai:2011it,Basu:2012fg}. The latter approach also provides a determination of the stellar age.

We performed a grid-based search using several available sets of evolutionary models to make an initial estimation of the stellar parameters of both targets. Details on how the sets of evolutionary tracks were constructed and the different search methods can be found in the following references: \citet{Basu:2010hv} and \citet{Gai:2011it} for the YY, YREC, Dotter and Marigo results; \citet{Stello:2009fg} for RADIUS; \citet{SilvaAguirre:2012du} for GARSTEC; \citet{Creevey:2012kc} for RadEx10; and \citet{Quirion:2010is} for SEEK. One grid-based pipeline built using the BaSTI\footnote{http://albione.oa-teramo.inaf.it/} isochrones \citep{Pietrinferni:2004im} is presented here for the first time.

Briefly, the BaSTI pipeline uses a denser set of isochrones in metallicity space than the publicly available ones, computed specially for the latest revision of the Geneva-Copenhagen Survey \citep{Casagrande:2011ji} and now extended for asteroseismic analysis. They are built assuming the \citet{Grevesse:1993vd} solar composition and adopting a value of $\Delta Y/\Delta Z=1.45$, in agreement with Big Bang nucleosynthesis (SBBN) values of primordial helium abundance at low metallicities \citep[$\mathrm{Y_P}=0.2482\pm0.0007$,][]{Steigman:2007ky,Steigman:2010gz}. The 22 available chemical mixtures cover a metallicity range between $-3.27<$[Fe/H]$<+0.5$. Models used here do not include overshooting and gravitational settling, and are calculated assuming mass loss according to the \citet{Reimers:1975vw} formulation with a fixed efficiency parameter of $\eta=0.4$. The algorithm to find the best match to the data is that of the GARSTEC grid, based on Monte Carlo realizations to form the probability distribution of each parameter.

Table~\ref{tab:grid} shows the results obtained with each grid pipeline for the two targets, including their 1-$\sigma$ uncertainties. As mentioned above, some methods return a probability distribution for the stellar parameters characterized by asymmetric error bars. The last column shows, for comparison, the values obtained with the direct method (Eqs.~\ref{eqn:mass}~and~\ref{eqn:rad}). There is an overall good level of agreement in the results for $\log\,g$ and $R/R_\odot$, explained by the fact that the global asteroseismic parameters are mostly sensitive to ratios between stellar mass and radius (mean density and surface gravity). Due to these strong dependencies, stellar radii and $\log\,g$ as determined from seismology are very precise \citep[][Huber et al. 2012]{Bedding:2011wl,Morel:2012de,SilvaAguirre:2012du}.

On the other hand, asteroseismic mass determinations still require verification from independent methods. The mean values from the grids are 1.08~$M_\odot$ for Perky and 1.23~$M_\odot$ for Dushera. All masses obtained by the 9 pipelines are contained within a 10\% range from these averages. In spite of that, when taking into account the 1-$\sigma$ uncertainties given by each grid, mass values between $\sim$0.95-1.20~$M_\odot$ and $\sim$1.05-1.35~$M_\odot$ are possible for Perky and Dushera, respectively. The large age differences in the results ($\sim$50\% or more) are a direct consequence of this mass difference ($\sim$0.3~$M_\odot$). Moreover, the fact that the results are in the vicinity of the $\sim$1.1~$M_\odot$ limit where the onset of core convection is thought to occur suggests that the existence and extent of a convective core will play a significant role in the internal structure of the star.

These results are to be taken with some caution since the scaling relations given in Eqs.~\ref{eqn:mass}~and~\ref{eqn:rad} are not exact. Moreover, they are of course dependent on the uncertainties in asteroseismic ($\langle\Delta\nu\rangle$ and $\nu_\mathrm{max}$) as well as atmospheric ($\teff$ and metallicity) inputs fed into the grid searches, which we decided to keep at a conservative level as explained in Section~\ref{s:targs}. In fact, each grid of models has been constructed using different sets of input physics, solar mixture, opacities, equation of state, etc., which accounts for some of the differences seen in the results (see Chaplin et al. (2013), in preparation, for a throughout discussion of the grid-to-grid scatter). When bearing this in mind it is clear that a detailed asteroseismic study is mandatory to determine masses with higher accuracy, to explore the energy transport mechanisms in the inner structure, and to better constrain the age of the targets. The lack of sensitivity of the global seismic analysis to the finer details of the stellar structure needs to be compensated with  tools capable of probing the deep interior of stars.
\section{Asteroseismic diagnostics for stellar interiors}\label{s:ratios}
Different combinations of oscillation frequencies have been proposed in the literature to study the internal structure of stars. The two most commonly used are the large frequency separation (Eq.~\ref{eq:large}) and the small frequency separation defined as
\begin{equation}\label{eq:small}
d_{\ell,\ell+2}(n)  =  \nu_{n,\ell}-\nu_{n-1,\ell+2}\,.
\end{equation}
The latter quantity is sensitive to the conditions in the stellar core and thus to the central hydrogen content (age) for main-sequence stars \citep[e.g.,][]{Ulrich:1986ge,1987Natur.326..257G,1989A&A...226..278G,Roxburgh:1994ux}. In contrast, the large frequency separation is known to be affected by the outer layers of stars \citep[see e.g.,][]{1988Natur.336..634C,ChristensenDalsgaard:1992ts,Roxburgh:2003bb,Roxburgh:2010ev}, where stellar models use simple versions of convection theory instead of properly dealing with turbulent pressure and non-adiabatic effects \citep[e.g.,][]{1992MNRAS.255..639B,1999A&A...351..689R,Ballot:2004dt,Houdek:2010ja}. For the solar case, differences in $\Delta\nu$ around the $\nu_{\mathrm{max},\odot}$ region between BiSON \citep[Birmingham Solar Oscillations Network,][]{1999MNRAS.308..424C} data and Model~S from \citet{ChristensenDalsgaard:1996gf} are of the order of 1$\mu$Hz, much larger than the typical solar observational uncertainties of $\sigma_{\Delta\nu}\sim$0.06$\mu$Hz.

In order to study the deep interior of the star we need combinations of frequencies that isolate the signal arising from the stellar center as much as possible from surface contamination. \citet{Roxburgh:2003bb} proposed to use ratios of small to large separations, which are mainly determined by the inner structure of the star \citep{Roxburgh:2003bb,OtiFloranes:2005ii,Roxburgh:2009kz}. They are constructed as
\begin{equation}\label{eq:r02}
r_{02}(n)=\frac{d_{0,2}(n)}{\Delta\nu_{1}(n)}
\end{equation}
\begin{eqnarray}\label{eq:rat}
r_{01}(n)=\frac{d_{01}(n)}{\Delta\nu_{1}(n)},& & r_{10} = \frac{d_{10}(n)}{\Delta\nu_{0}(n+1)}\,,
\end{eqnarray}
where $d_{01}(n)$ and $d_{10}(n)$ are the smooth 5-points small frequency separations:
\begin{equation}\label{eq:d01}
d_{01}(n)=\frac{1}{8}(\nu_{n-1,0}-4\nu_{n-1,1}+6\nu_{n,0}-4\nu_{n,1}+\nu_{n+1,0})
\end{equation}
\begin{equation}\label{eq:d10}
d_{10}(n)=-\frac{1}{8}(\nu_{n-1,1}-4\nu_{n,0}+6\nu_{n,1}-4\nu_{n+1,0}+\nu_{n+1,1})\,.
\end{equation}

The ratios of degree 0 and 1 (Eq.~\ref{eq:rat}) have already been used to identify the location of the convective envelope in the Sun \citep{Roxburgh:2009kz}, and also in other main-sequence targets \citep[e.g.,][]{Lebreton:2012hp,2012AN....333.1040M}. Inspection of Eqs.~\ref{eq:d01}--\ref{eq:d10} shows that $r_{01}(n)$ and $r_{10}(n)$ are alternatively centered in $\nu_{n,0}$ and $\nu_{n,1}$, respectively. Therefore we can consider the ratios $r_{01}(n)$ and $r_{10}(n)$ as one unique set of observables as a function of frequency which we call $r_{010}$:
\begin{equation}\label{eq:r010}
r_{010}=\{r_{01}(n),r_{10}(n),r_{01}(n+1),r_{10}(n+1),r_{01}(n+2),r_{10}(n+2),...\}\,.
\end{equation}

Existence of a growing convective core in stellar models produces an abrupt discontinuity in the chemical composition at the boundary between the core and the radiative envelope. The formal location of this boundary is determined by a stability criterion, most commonly the Schwarzschild criterion. When additional mixing such as overshooting is included in stellar models, the transition between the homogeneously mixed center and the radiative zone does not occur at the formal edge of the convective core but at a certain distance from it determined by the adopted mixing efficiency. In either case, the discontinuity in chemical composition (and thus adiabatic sound speed) produces a shift in the frequencies relative to those in a corresponding star with a smooth sound-speed profile that is an oscillatory function of the frequency itself, and whose period is related to the location of the discontinuity \citep[e.g.,][]{Roxburgh:1994p880,Roxburgh:2007eq,Cunha:2007gs}.

It is customary to represent the radial coordinate in stellar interiors by the wave travel time from the surface to the center, called the acoustic depth $\tau$, or its alternative representation named the acoustic radius:
\begin{eqnarray}\label{eq:acrad}
\tilde{t} & = & \int_{0}^{r}\frac{\mathrm{d}r}{c_s}\,,
\end{eqnarray}
where $c_s$ is the adiabatic sound speed. This quantity measures the travel time from the center towards the surface. Considering that the total acoustic radius is given by $\tilde{t}_\mathrm{tot} = \tilde{t}(R_{\rm tot})$, then $\tau = \tilde{t}_\mathrm{tot} - \tilde{t}$. If the sound-speed discontinuity is located at, say, $r_1$ in radial coordinates, and that same position is represented in acoustic radius and acoustic depth by $\tilde{t}_1$ and $\tau_1$ respectively, the periods of the oscillation in frequency space induced by this sharp variation are 1/(2$\tilde{t}_1$) and 1/(2$\tau_1$), according to the considered seismic coordinate \citep[e.g.,][]{2003MNRAS.344..657M}.

We stress the fact that the frequency combinations will be sensitive to the position of the transition between the homogeneously mixed and radiative region, regardless if this coincides or not with the formal boundary of the convective core. If the ratios were only affected by the location of the outer edge of the mixed region, information about its location could be directly extracted from the asteroseismic data. Nonetheless, the small acoustic radius characterizing the size of the convective core introduces an oscillatory signal with a period much larger than the observed frequency range. Moreover it has been shown that, although the ratios are mostly sensitive to the presence and size of a convective core, they are also affected by the central hydrogen content \citep{Brandao:2010fk,SilvaAguirre:2011jz} and still contain the signal of the base of the convective envelope \citep{Roxburgh:2003bb,Roxburgh:2010ev}.

There are several other ingredients of stellar evolution that play more or less important roles in the determination of the sound speed. Use of different opacities, equation of state, and convective theory applied affects the internal temperature stratification in stellar models. When using frequency ratios one must bear in mind that their behavior reflects the interplay between evolutionary stage (through the central hydrogen content) and a central convective region (when it is present), while the rest of the microphysics affects them to a much lesser (but not negligible and still not fully quantified) extent. A detailed study of these issues goes beyond the scope of this paper and will be presented elsewhere (Silva Aguirre et al. 2013, in preparation).

\begin{figure*}[!ht]
\centering
\includegraphics{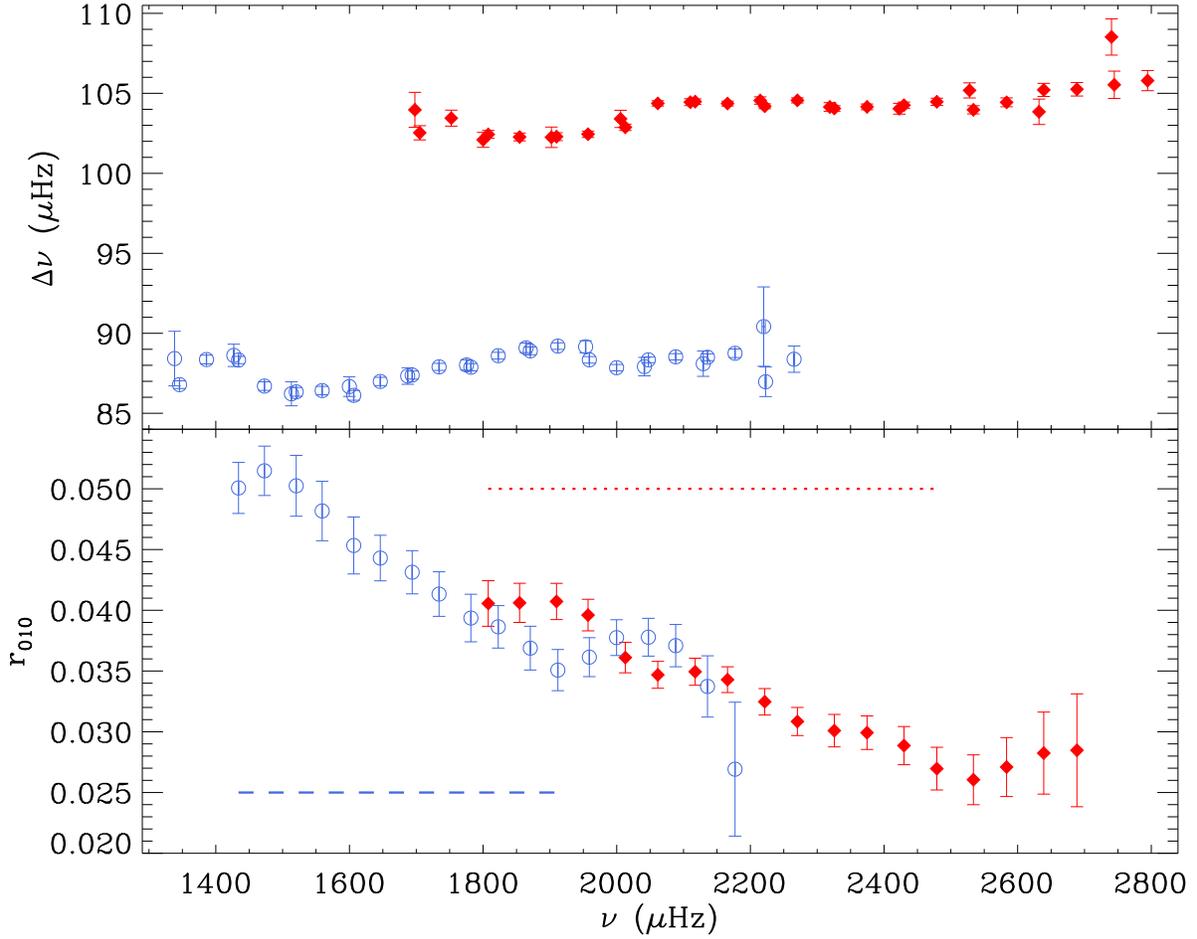}
\caption{Upper panel: large frequency separations $\Delta\nu_{0}, \Delta\nu_{1}, \Delta\nu_{2}$ of Perky (filled red diamonds) and Dushera (open blue circles). Lower panel: frequency ratios $r_{010}(n)$. The frequency range selected to fit the ratios is depicted by horizontal dotted and dashed lines for Perky and Dushera, respectively. See text for details.}
\label{fig:data}
\end{figure*}

Figure~\ref{fig:data} shows the observed large frequency separations and ratios $r_{010}$ for the two targets considered in this study. Inspection of the lower panel shows that the stars have different mean values and slopes in their ratios. Moreover, the oscillatory component introduced by the position of the convective envelope is clearly visible in the ratios of Perky.

The behavior of the frequency ratios in both targets shows a sudden increase at high frequencies, that occurs regardless of the length of the data used and is not predicted by models. We have compared individual frequencies extracted from datasets of different lengths and confirmed that, due to the low signal-to-noise ratio and the larger linewidth at the high-frequency end, this behavior is the result of realization noise. For this reason, we will not consider the frequency ratios on this regime when comparing to stellar models. In the lower panel of Fig.~\ref{fig:data} we mark the frequency range that will be taken into account to study $r_{010}$ \citep[the {\it linear range}, see][]{SilvaAguirre:2011jz}. The range considered to fit the ratios $r_{02}$ has also been restricted following the same criterion.

\begin{figure}[!ht]
\centering
\includegraphics[width=\linewidth]{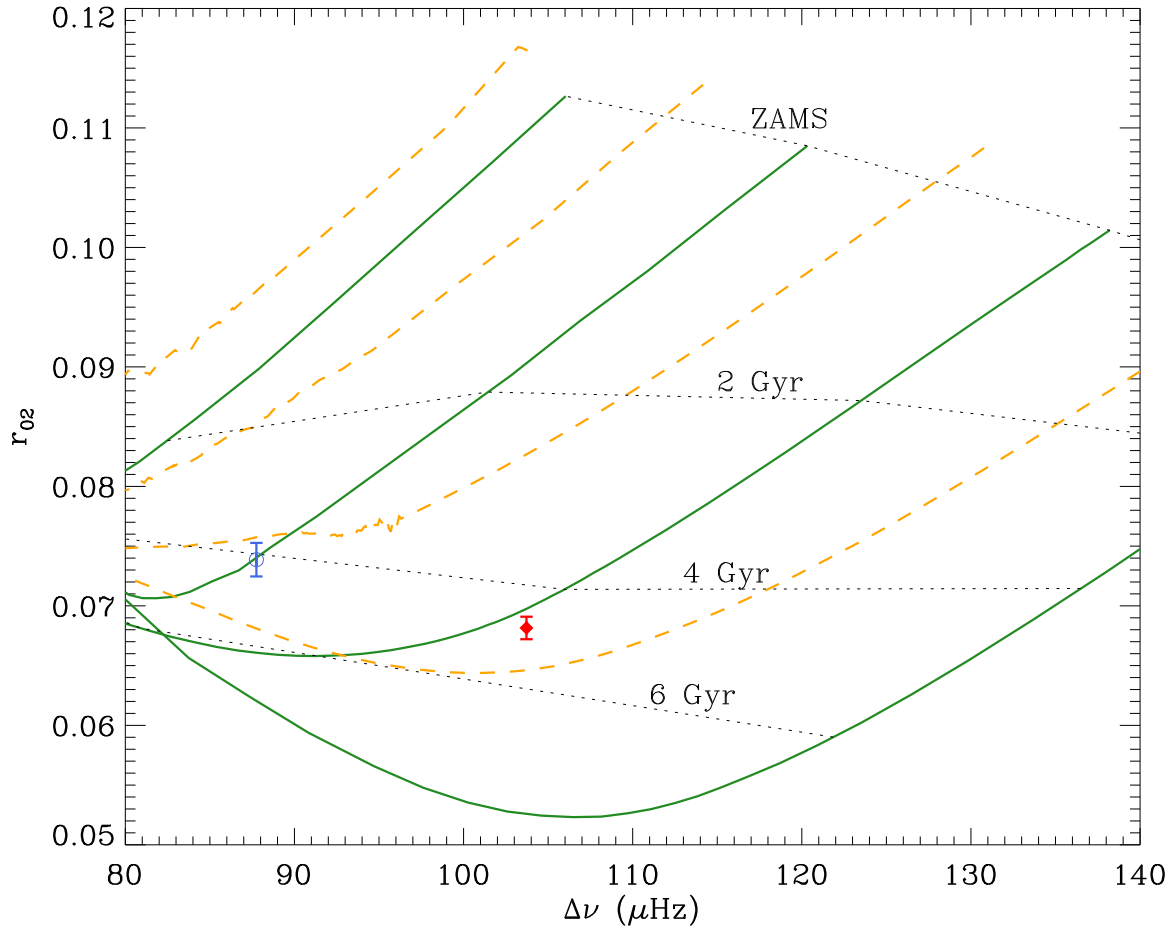}
\caption{Modified C-D diagram showing the location of Perky (filled red diamond) and Dushera (open blue circle). Evolutionary tracks are shown (from right to left) for masses 1.0-1.3~$M_\sun$ at solar ($Z=0.017$, solid lines) and sub-solar ($Z=0.011$, dashed lines) metallicities. Dotted lines connect positions at the same age for the solar metallicity tracks. Evolutionary tracks from \citet{White:2011fw}.}
\label{fig:data_r02}
\end{figure}

Together with $\langle\Delta\nu\rangle$ the average small frequency separation is used to construct an asteroseismic HR diagram known as the C-D diagram \citep{1984srps.conf...11C,ChristensenDalsgaard:1988wo}. A slightly modified version of this diagram is constructed by using the ratios $r_{02}$ instead of the small frequency separations, because it is a more effective indicator of stellar age \citep{White:2011fw}. In Fig.~\ref{fig:data_r02} we compare the position of Perky and Dushera in this diagram with evolutionary tracks at solar metalliticy, showing that for a given composition and input physics the age of the targets is well constrained by $r_{02}$. Also plotted in the figure are tracks with a metallicity $\sim$0.2~dex lower, showing that the offset due to composition has a strong impact in the derived ages from this diagram. This is particularly important for Dushera, as its position is closer to where the tracks collapse in this diagram when entering the subgiant phase of evolution.
%
\section{Detailed modeling}\label{s:modeling}
Asteroseismic modeling of individual targets has experienced a leap forward thanks to the accurate frequencies extracted from CoRoT and {\it Kepler} data \citep[e.g.,][]{Barban:2009gy,Ballot:2011jb,2011A&A...534A...6C,Appourchaux:2012kd}. The usual strategy applied in these type of studies consist on calculating different sets of one-dimensional hydrostatic stellar models. Using these as the equilibrium structures for a code capable of computing frequencies of oscillations, it is possible to produce a theoretical set of the spectroscopic and asteroseismic observables. Finally a criterion must be applied that defines which set of calculations best reproduces the available data \citep[e.g.,][]{2010ApJ...723.1583M,Brandao:2011dy,Mathur:2012bj,Dogan:2013fk}.

There are many ingredients in the computations of stellar models that are still not fully constrained, both in the microphysics (e.g., equation of state, opacities, nuclear reactions) and in the macrophysics (e.g., convective energy transport) of stellar evolution \citep[see e.g.,][]{2009AIPC.1111...55C,2013sse..book.....K}. Even when the same assumptions are used in different evolutionary codes, the resulting models can differ in their global properties, internal structure \citep[see e.g.,][]{Lebreton:2008fy,Montalban:2008ht}, and also in the theoretical frequencies obtained \citep{Moya:2008kb}.

For these reasons, several teams performed detailed modeling to match the stellar parameters presented in Table~\ref{tab:targ} and individual frequencies, using the stellar evolution and pulsation code of their choice (see Section~\ref{ss:evol} below). This allows us to obtain an at least qualitative measure of the systematic uncertainties inherent in the full modeling process, arising from the different input physics and implementations of search methodologies. A quantitative investigation of the individual influence of the various differences between the models presented here is beyond the scope of this paper.

Techniques to find the best-fit model to a given set of observables usually include minimization of a reduced $\chi^2$ \citep[e.g.,][]{Deheuvels:2011fn,Brandao:2011dy,Gruberbauer:2012cc} defined as
\begin{equation}\label{eq:chi2}
\chi^2 = \frac{1}{N} \sum_k \left( \frac{x_{\mathrm{obs}}(k) - x_{\rm model}(k)}{\sigma(x_{\rm obs}(k))} \right)^2 \,,
\end{equation}
where $k$ runs over the observable parameters, $x_{\rm obs}(k)$ represent each observable with its uncertainty $\sigma(x_{\rm obs}(k))$, $x_{\rm model}(k)$ are the theoretically produced values, and $N$ the total number of observations. In the analysis of the models that best fit the data, we will consider $\chi^2$ values for individual frequencies and ratios. It is clear from Eqs.~\ref{eq:d01}~and~\ref{eq:d10} that the differences entering the ratios $r_{010}$ are strongly correlated. We take this into account by including their covariance term in the $\chi^2$
\begin{equation}\label{eq:chi2_cov}
\chi^2 = \frac{1}{N} \left(\vec{x}_{\mathrm{obs}} - \vec{x}_{\rm model}\right)^{T} \, {\bf C}^{-1} \, \left(\vec{x}_{\mathrm{obs}} - \vec{x}_{\rm model}\right) \,,
\end{equation}
where ${\bf C}$ is the covariance matrix calculated after perturbing the observed frequencies using Monte Carlo simulations. To calculate the goodness of fit for the ratios $r_{010}$ we use Eq.~\ref{eq:chi2_cov}, which reduces to Eq.~\ref{eq:chi2} for the individual frequencies and ratios $r_{02}$ as they are independent.

As mentioned in Section~\ref{s:ratios}, theoretical frequencies of oscillations are not able properly to reproduce the observed ones due to our poor understanding of the physical processes taking place in the outer layers of stars \citep[see e.g.,][for the solar case]{ChristensenDalsgaard:1996gf}. There are different formulations to correct for this offset \citep[see e.g., the supplementary materials for][and references therein]{Carter:2012gq}, the most commonly used is that of \citet{Kjeldsen:2008kw}. It consists of an empirical power law derived using helioseismic data and standard solar models, which can be written as
\begin{equation}\label{eq:surf_cor}
\nu_\mathrm{obs}(n)-r\nu_\mathrm{ref}(n)=a\left(\frac{\nu_\mathrm{obs}(n)}{\nu_0}\right)^b\,.
\end{equation}
Here $\nu_\mathrm{obs}(n)$ correspond to the observed frequencies and $\nu_\mathrm{ref}(n)$ are the theoretical ones. To apply this surface correction one has initially to specify the values of the exponent $b$ and the reference frequency $\nu_0$. Then $r$ is determined after calculating an observed and theoretical average of the large frequency separation, obtained using a linear fit to the individual $\ell=0$ modes as a function of radial order $n$. The value of $a$ is determined from the average theoretical and observed $\ell=0$ frequencies \citep[see Eqs.~6,~8,~9~and~10 in][]{Kjeldsen:2008kw}. Although most modeling teams use this formulation in their analysis, they determine the parameters involved in the correction in slightly different manners (see Section~\ref{ss:surf_cor} below).

Another point where discrepancies can arise from is the estimation of the surface iron abundance in stellar models, to be compared with the [Fe/H] value determined from spectroscopic observations. Although the issue of the solar surface abundances is still a matter of debate \citep[see e.g.,][]{Grevesse:2010gq,Caffau:2010ik}, the observed [Fe/H] in stars barely depends on the considered solar reference because the largest discrepancies between the different solar mixtures come from the C, N, O and Ne elements \citep[e.g.,][]{Serenelli:2009ev}. On the other hand in stellar models it is assumed that
\begin{equation}\label{eq:sol_abund}
\log (Z/X) - \log(Z/X)_\sun \simeq \mathrm{[Fe/H]}\,,
\end{equation}
where $X$ and $Z$ are respectively the mass fraction of hydrogen and all elements heavier than helium, and $\log(Z/X)_\sun$ is the surface solar value. Equation~\ref{eq:sol_abund} clearly depends on the chosen solar mixture, and assumes that the fraction of each element comprising $Z$ is distributed according to the selected set of solar abundance ratios. The commonly adopted values of $(Z/X)_\sun$ vary for different mixtures between $0.0245$ \citep{Grevesse:1993vd} and $0.0181$ \citep{Asplund:2009eu}. This means that for models with the same $(Z/X)$ at the surface but computed with a different set of solar abundances, the theoretically determined [Fe/H] can vary by $\sim$0.13~dex, comparable to the uncertainties quoted in Table~\ref{tab:targ} from spectroscopic determinations.

Due to the issues aforementioned, and the large error bars in the spectroscopically determined [Fe/H] and $\teff$ (see Table~\ref{tab:targ}), we will not include these constraints in the calculations of goodness of fit via $\chi^2$. These observables were used by all teams as guidelines when exploring the parameter space (except for models~G, see Section~\ref{sss:modG} below), and will be considered as reference for comparison that only in extreme cases of discrepancy could be used to discard models.
\subsection{Evolutionary calculations and search algorithms}\label{ss:evol}
In the following sections we describe the evolutionary code, input physics, and fitting techniques adopted by each modeling team to find the best match for both targets.
\subsubsection{Models A}\label{sss:modA}
Models~A were computed using the Garching Stellar Evolution Code \citep[GARSTEC,][]{Weiss:2008jy}. We adopted the 2005 version of the OPAL equation of state \citep[EOS,][]{Rogers:1996iv,Rogers:2002cr} complemented by the MHD equation of state for low temperatures \citep{Hummer:1988kn}, low-temperature opacities from \citet{Ferguson:2005gn} and OPAL opacities for high temperatures \citep{Iglesias:1996dp}, the \citet{Grevesse:1998cy} solar mixture, and the NACRE \citep{Angulo:1999kp} compilation for thermonuclear reaction rates with the updated cross section for $^{14}\mathrm{N}(p,\gamma)^{15}\mathrm{O}$ from \citet{Formicola:2004dl}. Due to the large uncertainty in the surface composition of Dushera (see Table~\ref{tab:targ}), diffusion of helium and heavy elements was included only for the models of Perky using the \citet{Thoul:1994iz} prescription. The atmospheric stratification is given by Eddington's gray plane-parallel formalism.

Convective zones are treated with the mixing-length theory (MLT) as described by \citet{2013sse..book.....K}, using the solar calibrated value of the convective efficiency $\alpha_\mathrm{mlt}=1.791$. Overshooting is implemented in GARSTEC as a diffusive process consisting of an exponential decline of the convective velocities within the radiative zone \citep{Freytag:1996vw}. The diffusion constant is given by
\begin{equation}\label{eq:ove}
 D_{\mathrm{ov}}\left(z\right) = D_0 \ \exp \ \left(\frac{-2z}{\xi H_p}\right)\,,
\end{equation}
where $\xi$ corresponds to an efficiency parameter, $H_p$ is the pressure scale height, $z$ the distance from the convective border, and the diffusion constant $D_0$ is derived from MLT-convective velocities. Using open clusters, the calibrated value for the overshooting efficiency is defined as $\xi=0.016$. To prevent the survival of the pre-main sequence convective core due to overshooting in the solar model, a geometric factor for efficiency restriction is used when convective cores are small \citep{1999A&A...347..272S,Magic:2010iz}. When a convective core was present, we explored overshooting efficiencies between $\xi=0.004-0.035$.

Oscillation frequencies of the specific models analyzed were computed using the Aarhus Adiabatic Oscillation Package \citep[ADIPLS,][]{ChristensenDalsgaard:2008kr}. Models were selected by matching the observed large frequency separation and minimizing the $\chi^2$ fit to both frequency ratios $r_{010}$ and $r_{02}$. No surface correction was applied to define the best fit to the data.
\subsubsection{Models B}\label{sss:modB}
Models~B were computed using the Aarhus Stellar Evolution Code \citep[ASTEC,][]{ChristensenDalsgaard:2008bi}. The models are calculated using the NACRE compilation of nuclear reaction rates and the \citet{Grevesse:1993vd} solar mixture. The EOS is the OPAL 2005 version, while the opacities are those of OPAL for high temperatures and \citet{Ferguson:2005gn} for low temperatures. Convection is treated using the MLT as described by \citet{BohmVitense:1958vy}, and no microscopic diffusion is considered.

Core overshooting is implemented as an extension of the central mixed region using the radiative temperature gradient, and controlled by an efficiency parameter $\alpha_\mathrm{ov}$
\begin{equation}\label{eq:ove_adi}
d_\mathrm{ov}=\alpha_\mathrm{ov} \mathrm{min}(r_\mathrm{cc},H_p)\,,
\end{equation}
where $r_\mathrm{cc}$ is the radial size of the convective core. This prescription restricts overshooting if convective cores are smaller than $H_p$. Values of $\alpha_\mathrm{ov}=0.0, 0.1, 0.2$ were explored in the calculations of models~B, similar to the calibrated efficiencies of other codes with the same prescription (see models~F below).

The atmospheric stratification is given by the $T-\tau_{\mathrm{op}}$-relation of model~C in \citet{1981ApJS...45..635V}. Oscillation frequencies are computed using ADIPLS to find the best fit to the data by minimizing the $\chi^2$ values of individual frequencies and frequency ratios after applying the \citet{Kjeldsen:2008kw} surface correction. There is one subtle difference in the way the surface correction is applied in models~B, that will turn out to be of particular importance in the case of Perky (cf.~Eq.~\ref{eq:surf_cor}). To obtain the average large frequency separation needed to compute $r$, a fit is made to all available frequencies instead of only the radial modes. Similarly, radial and non-radial modes are used in the average frequencies used to obtain the $a$ parameter.
\subsubsection{Models C}\label{sss:modC}
Models~C were also computed using ASTEC with the same input physics as described for models~B, but no overshooting was included. A grid of models was calculated covering the following range in the mass-initial composition space: M$=1.0-1.6~M_\sun$, $Y=0.24-0.32$, and $Z/X=0.01-0.07$, while the mixing length parameter $\alpha_{\rm MLT}$ was fixed to 1.8. Frequencies were computed with ADIPLS for the models that have [Fe/H] and $\log\,g$ within 3-$\sigma$, and $\teff$ within 1-$\sigma$ from the values given in Table~\ref{tab:targ}, since the uncertainties on $\teff$ were relatively conservative. The best-fitting models were selected based on minimization of the $\chi^{2}$ value for only the individual frequencies after applying the \citet{Kjeldsen:2008kw} empirical surface correction.
\subsubsection{Models D}\label{sss:modD} 
Models~D were calculated using the CESAM2K evolutionary code \citep{1997A&AS..124..597M,Morel:2008kw}, with the OPAL 2005 EOS and opacities, the \citet{Grevesse:1993vd} solar mixture, and microscopic diffusion following \citet{1993ASPC...40..246M}. Convection is modeled using the \citet{1996ApJ...473..550C} formulation, which for this case gives a solar calibrated value for the convective efficiency of $\alpha_\mathrm{conv}=0.64$. Nuclear reactions are those from the NACRE compilation, while the atmosphere is described by Eddington's gray law and connected to the stellar interior at an optical depth of {\bf $\tau_{\mathrm{op}}=10$}. Core overshooting is implemented using an adiabatic temperature stratification and efficiency as given by Eq.~\ref{eq:ove_adi}.

Oscillation frequencies are computed using the Li\`{e}ge Oscillation code \citep[LOSC,][]{Scuflaire:2007fy}. The best-fit model is found via optimizations following the Levenberg-Marquardt algorithm, using spectroscopic constraints and matching individual frequencies after applying the \citet{Kjeldsen:2008kw} surface correction. The free parameters adjusted during the procedure are mass, age, initial helium abundance and metallicity, and the convective and overshooting efficiencies. The same procedure was used by \citet{Miglio:2005is} using the frequency ratios $r_{02}$ instead of individual frequencies.
\subsubsection{Models E}\label{sss:modE}
Models~E were computed using the Asteroseismic Modeling Portal \citep[AMP, e.g.,][]{Metcalfe:2012kv,Mathur:2012bj}. It is based on a genetic algorithm for minimization \citep{Metcalfe:2003ka} that matches the observed individual frequencies and spectroscopic constraints to those produced by a stellar evolution and pulsation codes. Its current implementation uses ASTEC for the evolutionary calculations, with similar input physics as models~B but low-temperature opacities from \citet{1994ApJ...437..879A} and nuclear reactions from \citet{1995RvMP...67..781B}. No overshooting is included.

Theoretical frequencies are computed with ADIPLS and corrected using the \citet{Kjeldsen:2008kw} formulation. The best-fit model is chosen by minimizing the sum of two separate $\chi^2$ values: one determined for the individual frequencies and another one for the spectroscopic constraints.
\subsubsection{Models F}\label{sss:modF}
Models~F were computed using the Yale Rotation and Evolution Code \citep{Demarque:2007ij}, in its non-rotating configuration. The code uses low-temperature opacities from \citet{Ferguson:2005gn} and OPAL opacities for high temperatures, OPAL EOS, the \citet{Grevesse:1998cy} solar mixture, and \citet{Adelberger:1998iv} nuclear reactions using the updated $^{14}\mathrm{N}(p,\gamma)^{15}\mathrm{O}$ cross section from \citet{Formicola:2004dl}. Models include helium and heavy-element diffusion \citep{Thoul:1994iz}, and overshooting as given by Eq.~\ref{eq:ove_adi} using a fixed value of $\alpha_\mathrm{ov}=0.2$ and an adiabatic temperature gradient when a convective core exists. This value of overshooting efficiency is motivated by the fit to the open cluster NGC~2420, and described in \citet{1994ApJ...426..165D}.

The atmospheric stratification is given by an Eddington $T-\tau$ relation, and frequencies calculated with the pulsation code used by \citet{1994A&AS..107..421A}. In order to find the best fit to the data, sequences of models were calculated with masses between 0.94-1.12~$M_\sun$ for Perky and 1.08-1.24~$M_\sun$ for Dushera. For each mass, we considered three different values of the mixing-length parameter, $\alpha_{\rm mlt}=$1.55,~1.7, and~1.826 (the solar calibrated value). The best fit was found using the reduced $\chi^2$ determined for the individual frequencies and frequency ratios assuming the solar surface term described in the supplementary material of \citet{Carter:2012gq}.
\subsubsection{Models G}\label{sss:modG}
Models G were constructed using ASTEC with the same input physics as models~B, exploring different values of overshooting efficiency $\alpha_\mathrm{ov}=0.0,0.1,0.2$ and metallicity $-0.5<\mathrm{[Fe/H]}<0.5$. No diffusion of elements was included in these calculations. A grid of models spanning a wide region of the Hertzsprung-Russell diagram was considered, with masses varying from 1.0~to~1.6 $M_\sun$. ADIPLS was used to compute the oscillation frequencies and the best models were selected solely based on their asteroseismic characteristics, in particular, their $r_{02}$ and $r_{010}$ ratios, the frequency derivatives (slopes) of the ratios and the large frequency separations. In short, the range of possible models was first reduced based on the following criteria: that the slopes of $r_{02}$ and $r_{010}$ were within the 1-$\sigma$ observed values and that the large frequency separation laid between the observed value $\pm$~3 $\mu$Hz. The identification of the best model within this subset was based on the minimum of the $\chi^2$ computed by comparing the ratios $r_{02}$ and $r_{010}$ with their observed values.
\subsubsection{Models H}\label{sss:modH}
Models~H were extracted from a grid of models calculated using GARSTEC, described in detail in \citet{SilvaAguirre:2012du}. Oscillation frequencies were computed using ADIPLS, and the best fits to the data were obtained looking for the minimum $\chi^2$ of the individual frequencies after applying the \citet{Kjeldsen:2008kw} surface correction. No further optimization beyond the pre-computed grid was made. Input physics is the same as for models~A with the following exceptions. The EOS is the FreeEOS \citep[A.~W.~Irwin\footnote{http://freeeos.sourceforge.net/}, see also][]{2003ApJ...588..862C}, hydrogen burning nuclear reaction rates are from \citet{Adelberger:2011fp}, and $\alpha_{\rm mlt}=1.811$ is fixed from the calibration of a standard solar model. Overshooting is implemented as in Eq.~\ref{eq:ove} with a fixed efficiency of $\xi=0.02$ (also constrained geometrically in the case of small convective cores and thin shells) and microscopic diffusion of helium and metals has been included in all these models.
\subsection{Applying the surface correction}\label{ss:surf_cor}
In the previous sections we showed that search strategies and criteria for defining which model best reproduces the data are not unique and vary among teams. Some consider only individual frequencies while others also take into account frequency combinations.

\begin{figure}[!ht]
\centering
\includegraphics[width=\linewidth]{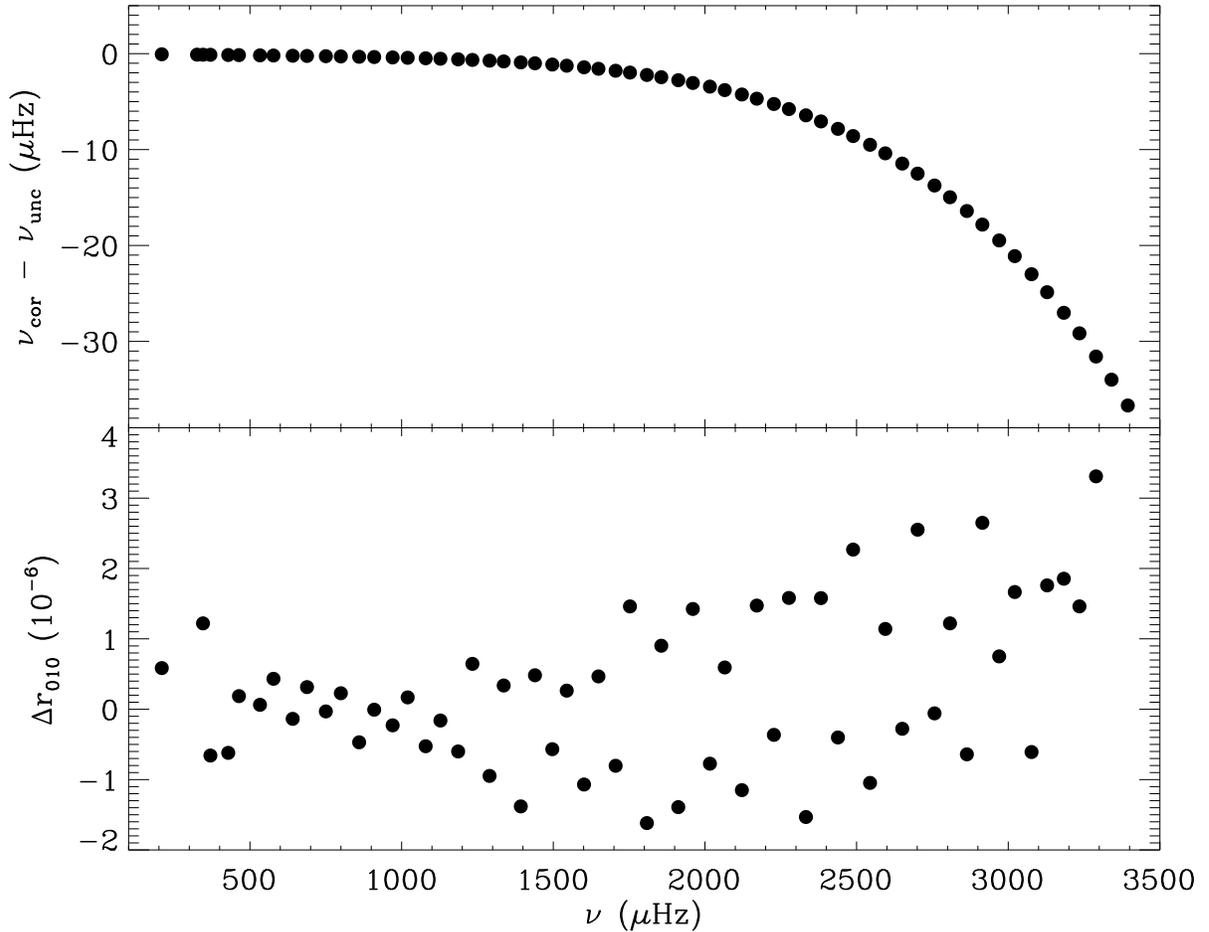}
\caption{Upper panel: frequency differences between corrected and uncorrected modes of oscillation. Typical uncertainties in the individual frequency determinations are of the order of $\sim$0.2$\mu$Hz. Lower panel: differences between corrected and uncorrected frequency ratios $r_{010}$. Typical uncertainties in the ratios are larger than the y-scale of the plot. See text for details.}
\label{fig:surf_eff}
\end{figure}

Figure~\ref{fig:surf_eff} depicts the effect of applying the surface correction in one of the models of Perky (cf. Eq~\ref{eq:surf_cor}). The upper panel shows the difference between corrected and uncorrected frequencies, presenting the same functional form as the solar case to which it is calibrated. The differences in the region of observed modes can be of up to $\sim20~\mu$Hz, a factor of 100 or more larger than the typical uncertainties in frequencies.

A different picture arises when looking at the difference between the ratios $r_{010}$ of corrected and uncorrected frequencies (lower panel in Fig.~\ref{fig:surf_eff}). Discrepancies are at the $10^{-6}$ level, three orders of magnitude below the average uncertainties, showing that this combination effectively cancels out the contribution from the outer layers \citep{Roxburgh:2003bb}. The value of the reduced $\chi^2$, when considering only individual frequencies, is thus heavily affected by the use of the surface correction, while the same is not true for the ratios.

Those teams applying the empirical surface correction of \citet{Kjeldsen:2008kw} consider somewhat different implementations of it. Before discussing how the parameters involved in the correction are actually obtained by each modeling team, a few words on the assumptions behind this formulation are necessary. It requires the value of $r$ to be close to unity, meaning that the mean stellar density of the model actually reproduces that of the real star. It considers only radial modes when finding the parameters for the corrections, assuming that non-radial frequencies are equally affected by the surface layers. Finally, usage of a single power law is justified by considering the offsets in the mode frequencies to be caused solely by the properties of the outer stellar layers, and thus dependent on surface gravity, effective temperature, and surface composition. As a consequence of these assumptions models found using this empirical treatment are expected to reproduce the global stellar properties (mass and radius), but not necessarily the core structure nor provide reliable age determinations \citep[see Section 5 in][]{Kjeldsen:2008kw}. Considering also non-radial modes in the fit is therefore necessary to ensure a proper representation of the deep stellar interior.

Obtaining the parameters involved in the surface correction deserves some comments. In principle, for each combination of evolutionary and pulsation code the power-law exponent $b$ should be determined from a solar calibration. This would also define the input physics to be used when modeling the targets, including the mixing-length parameter. In practice, variations of $b$ do not affect the obtained mean density of the models \citep{Kjeldsen:2008kw}, and one unique value can be used to model all stars. The $b$ parameters applied by the modeling teams in this study have been obtained either via a solar calibration of their own codes or taken from others. Therefore the adopted values vary among the results.

Another point where differences arise is in the calculation of the scaling parameter $r$. The best model for each star should have an $r$ value very close to unity. In the original formulation, this is determined from theoretical and observed frequencies of radial modes only. Although it is the most commonly adopted manner, some modeling teams do not consider this $r$-scaling, or they estimate $r$ using also the non-radial modes (models~B in Section~\ref{sss:modB} above). Finally, the reference frequency $\nu_0$ is arbitrarily chosen.

Since the surface correction is an empirical law based on the observed departure between the frequencies of the Sun and a solar model, there is no unique way of applying it. The net result is that $\chi^2$ values for individual frequencies determined by each team are not directly comparable due to these different implementations. Thus, we have uniformly applied the surface correction to all results to homogeneously compare them. As suggested by \citet{Kjeldsen:2008kw}, we fixed $b=4.9$, calculated $r$ only with $\ell=0$ modes, and considered the observed frequency of maximum power $\nu_{\mathrm{max}}$ from each star as the reference frequency $\nu_0$.
\subsection{Convective core location}\label{ss:ccore_loc}
\begin{figure}
\centering
\includegraphics[width=\linewidth]{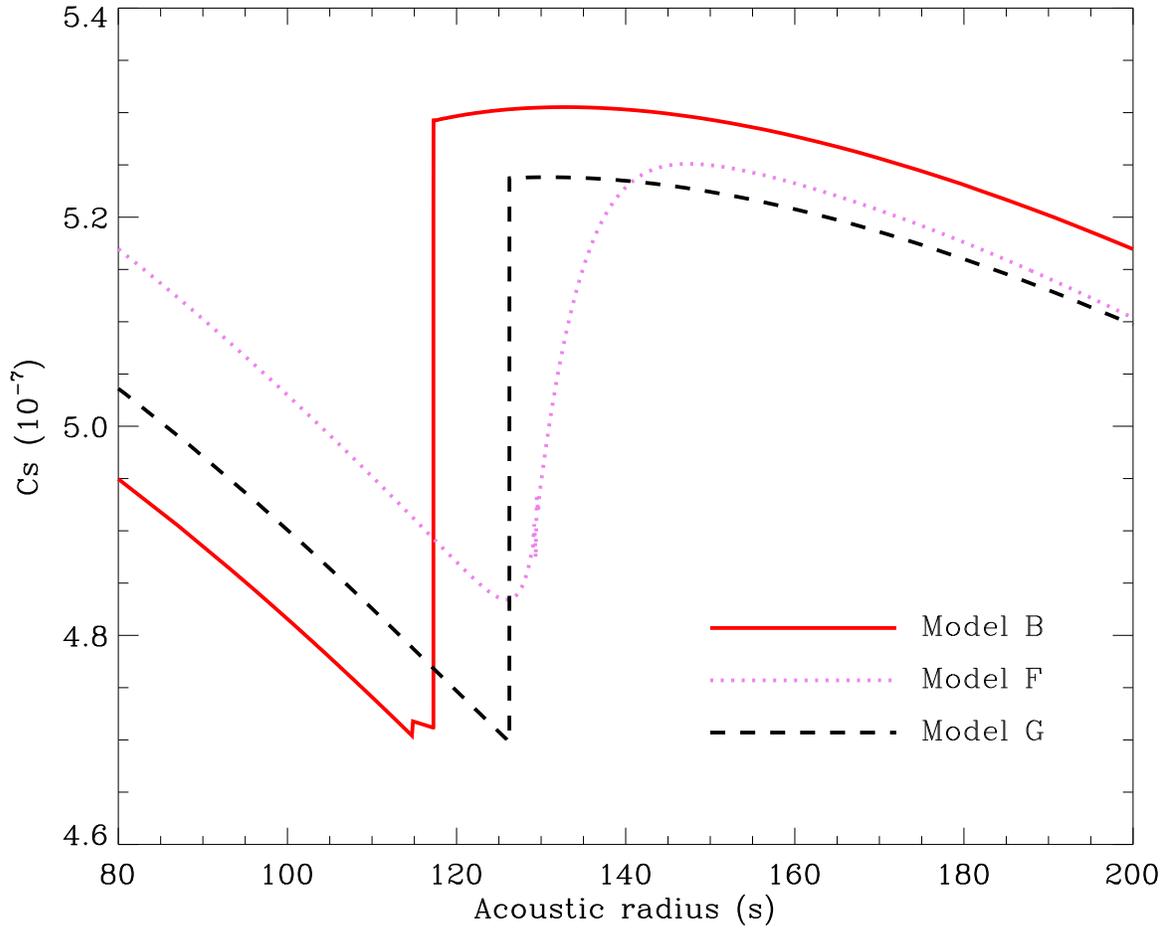}
\caption{Adiabatic sound speed near the stellar core as a function of acoustic radius for a set of selected models: B~(red solid line), F~(violet dotted line) , and G~(black dashed line).}
\label{fig:cs_dush}
\end{figure}

In Section~\ref{s:ratios} we described how the presence of a convective core produces an abrupt change in the adiabatic sound speed. The sharpness and shape of that transition will depend on the mixing processes taking place and the way overshooting is treated, if considered at all. In Fig.~\ref{fig:cs_dush} some examples of these transition regions are shown for models with convective cores, where distinctive features can be observed in different cases.

Model~B in Fig.~\ref{fig:cs_dush} does not include overshooting and shows simply a sharp vertical transition at the edge of the core. Model~G on the other hand applies a chemically homogeneous extension of the convective region when applying overshoot, and thus presents a similar structure beyond the formal limits of the convective core. In contrast model~F shows a smoother profile and a more extended transition region, produced by the interplay between overshooting and microscopic diffusion at the edge of the core.

Since different assumptions and evolutionary codes produce different sound-speed profiles, it is necessary to determine one criterion defining the position of the convective core for all models regardless of the considered input physics. As the p-modes propagate according to the sound speed, they are most sensitive to the regions where rapid variations in it take place. Therefore, we use the maximum in the adiabatic sound-speed derivative as a common definition for the boundary of the homogeneously mixed central region. This basically defines the extent of the mixed zone, which is equivalent to the formal convective core size unless overshooting is included.
\section{Results}\label{s:results}
In the following sections, we compare the best-fitting models found by each modeling team as described in Section~\ref{ss:evol}. It is worth mentioning that, when comparing radius determinations, we decided to use the radius measured at the photosphere in order to avoid systematics introduced by the different prescription for atmospheric stratification used in each evolutionary code.

In all cases, the values of $\chi^2_\nu$ for frequencies and $\chi^2_{r02}$~and~$\chi^2_{r010}$ for frequency ratios (Eqs.~\ref{eq:r02}~and~\ref{eq:rat}) given in the results below are calculated using frequencies corrected as explained in Section~\ref{ss:surf_cor}. For comparison purposes we also give the results of the fit to the individual frequencies using the original frequencies as corrected by each modeling team, which we denote $\chi^2_{\nu\, \mathrm{orig}}$. Since the frequency correction does not affect the computed ratios, we will sort models according to the average between $\chi^2_{r010}$~and~$\chi^2_{r02}$. We remind the reader that the correlations in the ratios $r_{010}$ are taken into account when computing $\chi^2_{r010}$, as mentioned in Section~\ref{s:modeling}.
\subsection{Perky}\label{ss:perk}
\begin{table*}[!ht]\scriptsize
\caption{Main characteristics of the models calculated for Perky from each method as described in Section~\ref{ss:evol}.}
\label{tab:model_perk}
\centering
\begin{tabular}{c c c c c c c c c c}
\hline\hline
\noalign{\smallskip}
Model & M ($M_\odot$) & Age (Gyr) & $\log\,g$  & $T_\mathrm{eff}$ (K) & $R_\mathrm{Phot}/R_\odot$ & [Fe/H]\footnote{Using the \citet{Grevesse:1993vd} solar mixture, see Eq.~\ref{eq:sol_abund}} & $Z_i$ & $Y_i$ & $\log\,(L/L_\odot)$\\ 
\noalign{\smallskip}
\hline
\noalign{\smallskip}
Model A & 1.09 & 4.62 & 4.299 & 6074 & 1.2257  & -0.09 &  0.0181  & 0.273  & 0.263 \\ 
\noalign{\smallskip}
Model B & 1.11 & 4.36 & 4.305 & 6122 & 1.2284 & +0.00 & 0.0175 & 0.267 & 0.279\\ 
\noalign{\smallskip}
Model C & 1.05 & 5.53 & 4.294 & 5970 &  1.2093 & -0.09 & 0.0145  & 0.260 & 0.222\\ 
\noalign{\smallskip}
Model D & 1.17 & 4.97 & 4.306 & 6016 & 1.2570 & -0.01 &  0.0210 & 0.240 & 0.268\\ 
\noalign{\smallskip}
Model E & 1.17 & 4.94 & 4.307 & 5951 &  1.2580 & +0.01 & 0.0197  & 0.226 & 0.250\\ 
\noalign{\smallskip}
Model F & 1.12 & 4.88 & 4.301 & 6100 &  1.2370 & -0.09 &  0.0179 & 0.264 &  0.279 \\ 
\noalign{\smallskip}
Model G & 1.06 & 4.47 & 4.299 & 6071 & 1.2088  & -0.10 &  0.0139 & 0.271 & 0.250\\ 
\noalign{\smallskip}
Model H & 1.13 & 3.80 & 4.305 & 6134 & 1.2382 & -0.07 &  0.0188 & 0.269  & 0.289 \\ 
\noalign{\smallskip}
\hline
\end{tabular}
\end{table*}
\begin{table*}[!ht]\scriptsize
\caption{Structure and seismic characteristics of the models for Perky. Values of $\chi^2_{\nu\, \mathrm{orig}}$ were obtained using the frequencies as corrected by each modeling team, while $\chi^2_{\nu}$ was calculated using the standard version of the surface correction. Last column gives the symbol used to plot each model in all figures.}
\label{tab:ast_perk}
\centering
\begin{tabular}{c c c c c c c c c c c}
\hline\hline
\noalign{\smallskip}
Model & $X_c$ & $\tilde{t}_\mathrm{cc}$ (s)& $\tilde{t}_\mathrm{Phot}$ (s)& $\tilde{t}_\mathrm{Tot}$ (s)& $\tilde{t}_\mathrm{cc}/\tilde{t}_\mathrm{Phot}$ (\%) & $\chi^2_{\nu\, \mathrm{orig}}$ & $\chi^2_{\nu}$ & $\chi^2_{r010}$ & $\chi^2_{r02}$ & Symbol \\
\noalign{\smallskip}
\hline
\noalign{\smallskip}
Model A & 0.124  & --  &  4549.5 & 4577.4 & -- & -- & 21.28 & 0.79 & 0.35 & Filled circle \\
\noalign{\smallskip}
Model B & 0.249  &  95.0 & 4541.9 & 4621.7 & 2.09 & 9.15 & 8.96 & 1.07 & 0.43 & Open square \\
\noalign{\smallskip}
Model C  & 0.078  & --  & 4557.5 & 4593.7 & -- & 11.03 &15.40 & 1.52 & 1.12 & Filled cross \\
\noalign{\smallskip}
Model D & 0.085  & --  & 4565.3 & 4691.6 & -- & 2.37 & 3.25 & 1.67 & 1.21 & Filled square \\
\noalign{\smallskip}
Model E  & 0.120  & --  & 4573.0 & 4633.3 & -- & 5.04 & 6.05 & 2.23 & 1.39 & Downward triangle \\
\noalign{\smallskip}
Model F & 0.085  & --  &   4553.4 & 4725.7 & -- & 3.50 & 10.27 & 1.04 & 3.62 & Open diamond \\
\noalign{\smallskip}
Model G  & 0.122  & --  &   4546.3 & 4626.1 & -- & -- & 21.78 & 2.22 & 3.56 & Upward triangle \\
\noalign{\smallskip}
Model H & 0.195 & --  & 4547.6 & 4613.6 & -- & 19.19 & 20.13 & 1.97 & 5.46 & Filled diamond \\
\noalign{\smallskip}
\hline
\end{tabular}
\end{table*}
The main characteristics of the best-fit model found by each team for the target Perky are given in Table~\ref{tab:model_perk}. The first feature to notice is that the agreement in surface gravity between all models is better than $0.02$~dex. This level of precision is already better than what was found using the direct and grid-based methods (see Table~\ref{tab:grid}), and the obtained mass range from detailed modeling is now restricted to values between 1.05-1.17~$M_\odot$. This decrease is partly responsible for the reduction in the age difference compared to the grid results, now ranging from $\sim$3.8 to $\sim$5.5 Gyr. The age spread is constrained within a level of $\sim$18\%, which although not extremely good is much better than that obtained from the global asteroseismic fit (see Section~\ref{s:dir_and_grid}). Effective temperatures are also contained within the values given in Table~\ref{tab:targ}. The overall agreement in radius is of the order of 0.05~$R_\odot$, only slightly better than that found with the global analysis.

Models~D~and~E have initial helium abundances below the currently accepted SBBN value. This problem occurs regularly when models are selected solely on the base of best reproducing the observed individual frequencies after applying the surface correction \citep[e.g.,][]{2010ApJ...723.1583M,Mathur:2012bj}. Interestingly enough, models~D~and~E have basically the same characteristics. Both cases present the highest mass values among the results, hinting that the correlation between mass and helium abundance through the mean molecular weight is reflected in the asteroseismic fits \citep[see also Fig.~5 in][]{Metcalfe:2009ed}. The initial metallicities of these models are also the highest ones, probably to compensate for the low helium abundance when achieving a similar luminosity and effective temperature. This might point towards deficits in the use of opacity or EOS data. We discuss this further in Section~\ref{s:disc}.

As we aim to make a better characterization of the target properties we analyze the spectrum of oscillations in detail. In Table~\ref{tab:ast_perk} the internal structure and asteroseismic properties are given for the best-fit models of Perky. The values of $\chi^2_{r010}$ have been calculated using the frequency range marked in Fig.~\ref{fig:data}. Interestingly, the models with the lowest $\chi^2_{\nu}$ value for the individual frequencies are not the ones that best reproduce the observed ratios, and are in fact those with the lower than SBBN initial helium abundances (D~and~E). Model~F presents a large difference between the original surface correction $\chi^2_{\nu\, \mathrm{orig}}$ and the one we applied ($\chi^2_{\nu}$). This model was the only one selected using a prescription different from the \citet{Kjeldsen:2008kw} one (see Section~\ref{ss:surf_cor}), reflecting the large impact of the assumed recipe for surface correction and warning us about selecting models solely based on fits to individual frequencies.

\begin{figure*}[!ht]
\centering
\includegraphics[width=\linewidth]{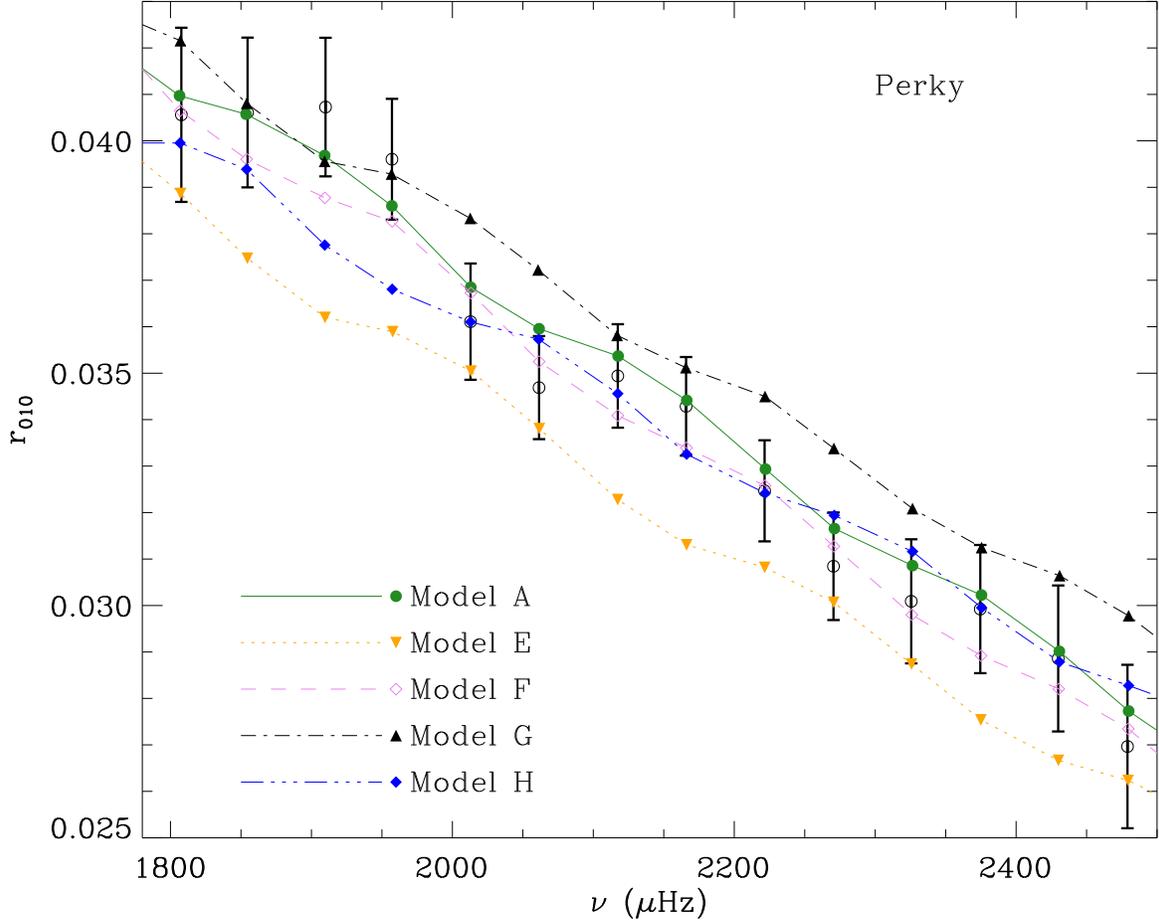}
\caption{Frequency ratios $r_{010}(n)$ for a selected set of models for Perky, and data only in the range marked in Fig.~\ref{fig:data}. Observational frequencies of Perky are depicted in open circles, while models~A,~E,~F,~G and~H are shown as filled circles, downward triangles, open diamonds, upwards triangles, and filled diamonds, respectively.}
\label{fig:rat_perk}
\end{figure*}

\begin{figure}[!ht]
\centering
\includegraphics[width=\linewidth]{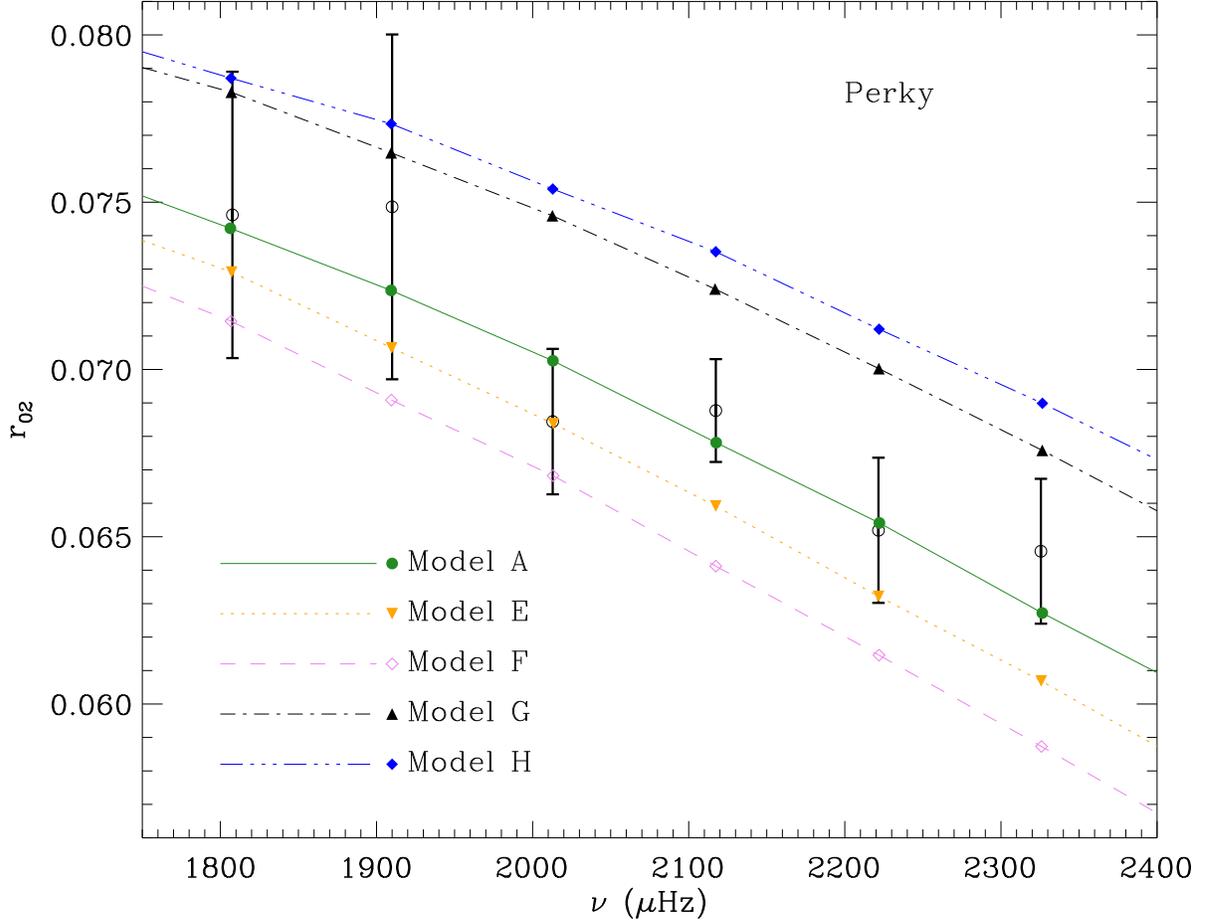}
\caption{Frequency ratios $r_{02}$ for the same models of Perky as in Fig.~\ref{fig:rat_perk}. Data for only those modes considered in the fit are plotted (see Sect.~\ref{s:ratios}).}
\label{fig:rat02_perk}
\end{figure}

A more detailed picture can be made when looking at Figures~\ref{fig:rat_perk} and~\ref{fig:rat02_perk}, where we show the comparison to the data of models~A,~E,~F,~G and~H. From Fig.~\ref{fig:rat_perk} it can be seen that the ratios $r_{010}$ of model~E~(G) are systematically below (above) the observed ones, allowing us to discriminate against them. Ratios $r_{02}$ in Fig.~\ref{fig:rat02_perk} have larger error bars and less data, but those of model~G are also in disagreement with the observations. Following the same line of reasoning models~F and~H can also be discarded due to their large discrepancy with the observed values of $r_{02}$.

More interesting than discarding models is trying to understand the structural differences that produce this behavior. As discussed in Section~\ref{s:ratios}, the ratios $r_{02}$ are a proxy for age and thus central hydrogen content of the star. Inspection of Table~\ref{tab:ast_perk} does not reveal a correlation between $X_c$ and the goodness of fit to $r_{02}$. In fact, models with the same central hydrogen content (D~and~F) present very different values of $\chi^2_{r02}$, while model~B gives an excellent fit to these ratios with more than twice the amount of hydrogen remaining in the core than any other model. Because of the systematic differences in the input physics and evolutionary codes, we no longer find a correlation between $r_{02}$ and $X_c$.

The case of model~B presents an interesting challenge to our current understanding of asteroseismic diagnosis. With the exception of this one, all other results point towards a star with no convective core. Nonetheless, this model was selected using a modified version of the \citet{Kjeldsen:2008kw} surface correction that includes also the non-radial modes (see Section~\ref{ss:surf_cor}), naturally resulting in a $\chi^2_{\nu}$ value that is slightly different than what would be obtained using the radial modes only. Interestingly enough, the algorithm applied to select model~B returns a different best-fit case when the "normal" surface correction is applied (using only $\ell=0$ modes). This new solution, which we call model~B', gives values of $\chi^2_{\nu\, \mathrm{orig}}=9.27$ and $\chi^2_{\nu}=8.18$. When compared to the results of model~B given in Table~\ref{tab:ast_perk} of $\chi^2_{\nu\, \mathrm{orig}}=9.15$ and $\chi^2_{\nu}=8.96$ we see that model~B' was not selected because it used the modified surface correction, while it would have been chosen if the standard one was applied instead.

Model~B', has exactly the same initial composition as model~B in Table~\ref{tab:model_perk}, but with a mass of 1.10~$M_\sun$ and no convective core present. Its age is 4.72~Gyr, $\teff=$6105~K, $\log\,g=4.300$ and luminosity $\log\,(L/L_\odot)=0.275$, in good agreement with the results of model~A and similar to model~B. However its central hydrogen abundance of $X_c=0.03$ is much lower than found in the other solutions, and gives fits to the frequency ratios of $\chi^2_{r010}=3.55$ and $\chi^2_{r02}=0.43$.

\begin{figure*}[!ht]
\centering
\includegraphics[width=\linewidth]{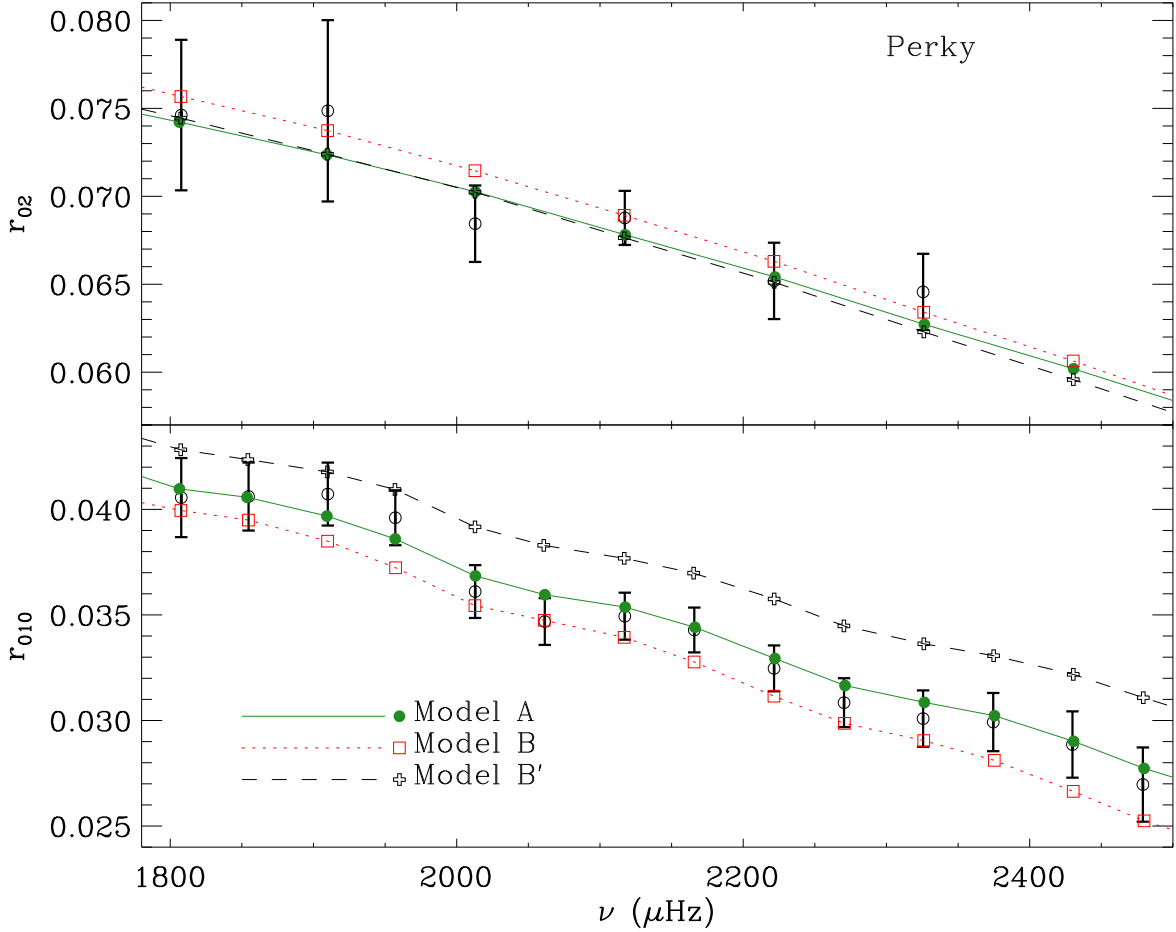}
\caption{Upper panel: frequency ratios $r_{02}$ of Perky. Lower panel: frequency ratios $r_{010}(n)$ of Perky. Observational frequencies are depicted by open circles, while models~A,~B, and~B' are shown as filled circles, open squares, and open crosses, respectively.}
\label{fig:rat_perkJCD}
\end{figure*}

Figure~\ref{fig:rat_perkJCD} shows the comparison to the data of models~A,~B, and~B'. Despite the very different central hydrogen contents (0.115, 0.249 and 0.03 for models~A,~B, and ~B', respectively), the ratios $r_{02}$ show no significant sensitivity to this parameter. Inspection of the ratios $r_{010}$ allows us to discriminate against model~B', but it is clear that for this particular case our current diagnostic tools are not capable of discriminating against the presence of a convective core (i.e. between models~A~and~B). We remind the reader that using the standard formulation of the surface correction from Eq.~\ref{eq:surf_cor} would have resulted in model~B' as the best fit found by this team, a model clearly not compatible with the observed ratios $r_{010}$.

To explore this issue in more detail, we reproduced the results of model~B with the evolutionary code and input physics of model~A. For the same initial metallicity and mass it was possible to sustain a convective core beyond the pre-main sequence phase only by removing the geometrical restriction for small convective cores (see Section~\ref{sss:modA}). The resulting model has an age of 4.71~Gyr at $\teff=6148$~K, $\log\,g=4.303$ and luminosity $\log\,(L/L_\odot)=0.288$. Convective core size is slightly smaller than the one found in model~B, placed at an acoustic radius of $\tilde{t}=81.2$~s that corresponds to 1.77\% of the total photospheric acoustic radius. The fit to the frequency data is also comparably good as in model~B, with $\chi^2_{r010}=1.45$ and $\chi^2_{r02}=0.32$.

The overall picture we are left with is the following. Based on the analysis described in the previous paragraphs, we can discard models~E,~F,~G, and~H and obtain from the rest of the results a mean and standard deviation pointing towards a star of $R=1.23\pm0.02~R_\sun$, $M=1.11\pm~0.05~M_\sun$, and $4.87\pm0.50$~Gyr. This corresponds to uncertainties of 1.6\%, 4.5\% and 10.4\% in radius, mass, and age, respectively; the level of precision obtained in the stellar parameters by this method of asteroseismic analysis is extremely good, particularly in age. However the experience of model~B shows that slightly different models with equally good or even better fits to the data can still be lurking among our results, and have not been picked out due to our assumptions on the fitting algorithms. Although the mass, radius and age are not particularly affected by this issue, there could be other sources of systematic errors arising from unexplored input physics. In any case, we expect the true uncertainties to be only sightly larger given the heterogeneous set of codes and input physics used in this analysis.

We are left with the unsolved issue of whether a convective core exists or not and, in relation to that, what the amount of hydrogen still remaining for central burning is. This has a direct impact in the inferred age at the turn-off: while model~A will remain in the main sequence for $\sim$1~Gyr, model~B will spend another $\sim$2~Gyr in this evolutionary phase.
\subsection{Dushera}\label{ss:dush}
\begin{table*}[!ht]\scriptsize
\caption{Main model characteristics of the best fits to the data of Dushera. Models are sorted according to the value of their $\chi^2_{r010}$ as given in Table~\ref{tab:ast_dush}. Last column gives the value of the overshooting efficiency parameter used.}
\label{tab:model_dush}
\centering
\begin{tabular}{c c c c c c c c c c}
\hline\hline
\noalign{\smallskip}
Model & M ($M_\odot$) & Age (Gyr) & $\log\,g$  & $T_\mathrm{eff}$ (K) & $R_\mathrm{Phot}/R_\odot$ & [Fe/H]\footnote{Using the \citet{Grevesse:1993vd} solar mixture, see Eq.~\ref{eq:sol_abund}} & $Z_i$ & $Y_i$ & OV\\
\noalign{\smallskip}
\hline
\noalign{\smallskip}
Model F & 1.12 & 3.53 & 4.208 & 6318 & 1.3803 & -0.15 & 0.0162 & 0.310 & $\alpha_\mathrm{ov}=$ 0.2\\ 
\noalign{\smallskip}
Model A & 1.17 & 4.06 & 4.215 & 6305 & 1.3990 & -0.12 & 0.0135 & 0.261 & $\xi=$ 0.03 \\ 
\noalign{\smallskip}
Model D & 1.26 & 3.09 & 4.223 & 6305 & 1.4378 & -0.03 & 0.0202 & 0.266 &  $\alpha_\mathrm{ov}=$ 0.015\\ 
\noalign{\smallskip}
Model E & 1.19 & 3.44 & 4.217 & 6262 & 1.4083  & -0.11 & 0.0147  & 0.263 &  $\alpha_\mathrm{ov}=$ 0 \\ 
\noalign{\smallskip}
Model C & 1.20 & 3.34 & 4.218 & 6094 & 1.4119 & +0.09 & 0.0210 & 0.280 & $\alpha_\mathrm{ov}=$ 0 \\ 
\noalign{\smallskip}
Model G & 1.14 & 4.28 & 4.208 & 6040 & 1.3926 & +0.00 & 0.0173  & 0.278 & $\alpha_\mathrm{ov}=$ 0.1\\ 
\noalign{\smallskip}
Model B & 1.20 & 3.67 & 4.216 & 6205 & 1.4157  & +0.00 & 0.0175  & 0.267 &  $\alpha_\mathrm{ov}=$ 0 \\ 
\noalign{\smallskip}
Model H & 1.19 & 4.00 & 4.214 & 6150 & 1.4119 & -0.08 & 0.0188 & 0.269 &  $\xi=$ 0.02 \\ 
\noalign{\smallskip}
\hline
\end{tabular}
\end{table*}
\begin{table*}[!ht]\scriptsize
\caption{Internal structure and seismic characteristics for the models of Dushera. Values of $\chi^2_{\nu\, \mathrm{orig}}$ were obtained using the frequencies as corrected by each modeling team, while $\chi^2_{\nu}$ was calculated using the standard version of the surface correction.}
\label{tab:ast_dush}
\centering
\begin{tabular}{c c c c c c c c c c c c}
\hline\hline
\noalign{\smallskip}
Model & $X_c$ & $\tilde{t}_\mathrm{cc}$ (s) & $M_\mathrm{cc}$ (\%) & $R_\mathrm{cc}$ (\%) & $\tilde{t}_\mathrm{Phot}$ (s) & $\tilde{t}_\mathrm{Tot}$ (s) & $\tilde{t}_\mathrm{cc}/\tilde{t}_\mathrm{Phot}$ (\%) & $\chi^2_{\nu\, \mathrm{orig}}$ & $\chi^2_{\nu}$ & $\chi^2_{r010}$ & $\chi^2_{r02}$\\
\noalign{\smallskip}
\hline
\noalign{\smallskip}
Model F  &  0.236 & 129.5  & 8.32 & 7.10 & 5367.1 & 5577.5 & 2.41 & 7.09 & 11.81 & 0.92 & 0.93 \\
\noalign{\smallskip}
Model A &  0.262 & 131.4 & 7.94 & 7.08 & 5368.1 & 5401.2 & 2.45 & -- & 20.3 & 1.44 & 0.47 \\
\noalign{\smallskip}
Model D &  0.256 &  125.1 &  5.89 & 6.57 &5387.4 & 5546.1 & 2.32 & 2.04 & 1.90 & 1.60 & 0.53\\
\noalign{\smallskip}
Model E  &  0.213 & 110.1  &  4.77 & 5.79 & 5276.0 & 5347.2 & 2.09 & 4.78 & 4.67 & 1.74 & 0.72\\
\noalign{\smallskip}
Model C  & 0.236 & 116.0 & 5.01 & 6.09 & 5390.8 & 5433.9 & 2.15 & 9.93 & 10.31 & 2.06  & 0.64 \\
\noalign{\smallskip}
Model G   &  0.214 & 126.2  &  6.94 & 6.63 & 5396.9 & 5498.1  & 2.34 & -- & 9.16 & 0.81 & 2.91\\
\noalign{\smallskip}
Model B &  0.173 &  117.3  &  5.28 & 5.99 & 5319.5 & 5417.1 & 2.21 & 4.72 & 3.87 & 1.58 & 2.30\\
\noalign{\smallskip}
Model H  & 0.243 &  122.0  &  6.66 & 6.67 & 5373.7 & 5430.9 & 2.27 & 20.98 &16.93 & 1.92 & 4.10\\
\noalign{\smallskip}
\hline
\end{tabular}
\end{table*}
For our second target of interest, Dushera, we performed a similar analysis to the one made for Perky. In Table~\ref{tab:model_dush} we give the main characteristics of the best-fit model found by each team, ranked again by their average fit to the frequency ratios. In this case no model gives a result with an initial helium abundance below the SBBN value. Agreement in $\log\,g$ values is again better than 0.02~dex, while the radii of the results are contained within a range of 0.06~$R_\sun$.

The mass range of the models is similar to the one obtained via the grid-based method, but with an age spread of the order of $\sim15$\%. All analyses for this target returned models where a convective core exists, but there are differences in the size of the centrally mixed region that we aim to characterize using the frequency combinations. Table~\ref{tab:ast_dush} presents the internal structure characteristics of each best-fit model, as well as the values of the fits to individual frequencies and ratios.

Once again, inspection of Table~\ref{tab:ast_dush} reveals that models best fitting the individual frequencies after surface correction (D and~B) do not necessarily give the best match to the frequency combinations. Just as in the case of Perky, the data warn us to be cautious when relying in results solely based on fits to individual frequencies. Also, comparison of the values given in Table~\ref{tab:ast_dush} shows no significant correlation between the central hydrogen content $X_c$ and $r_{02}$ among these models built with different input physics.

\begin{figure*}[!ht]
\centering
\includegraphics[width=\linewidth]{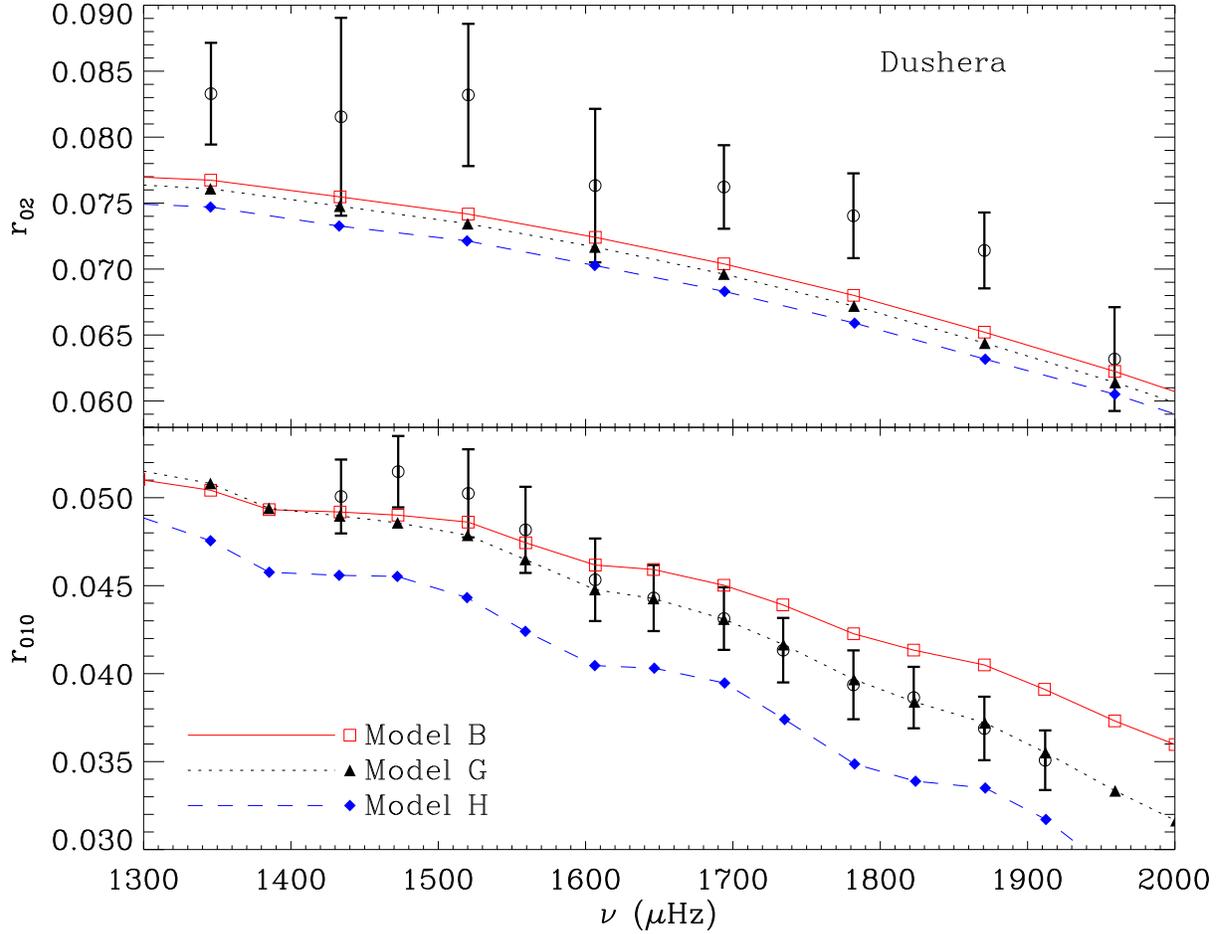}
\caption{Same as Fig.~\ref{fig:rat_perkJCD} for Dushera. Shown models are B~(open squares), G~(filled upward triangles), and H~(filled diamonds).}
\label{fig:ratW_dush}
\end{figure*}

We can further constrain the properties of the target by analyzing the fits to the ratios, plotted in Fig.~\ref{fig:ratW_dush} for the three models presenting the largest average of $\chi^2_{r010}$ and $\chi^2_{r02}$. Models~H is systematically below the observed ratios $r_{010}$, while model~B shows a slope that is not consistent with the data. Moreover, when looking at the ratios $r_{02}$, all models are systematically lower than the observed values.

\begin{figure*}[!ht]
\centering
\includegraphics[width=\linewidth]{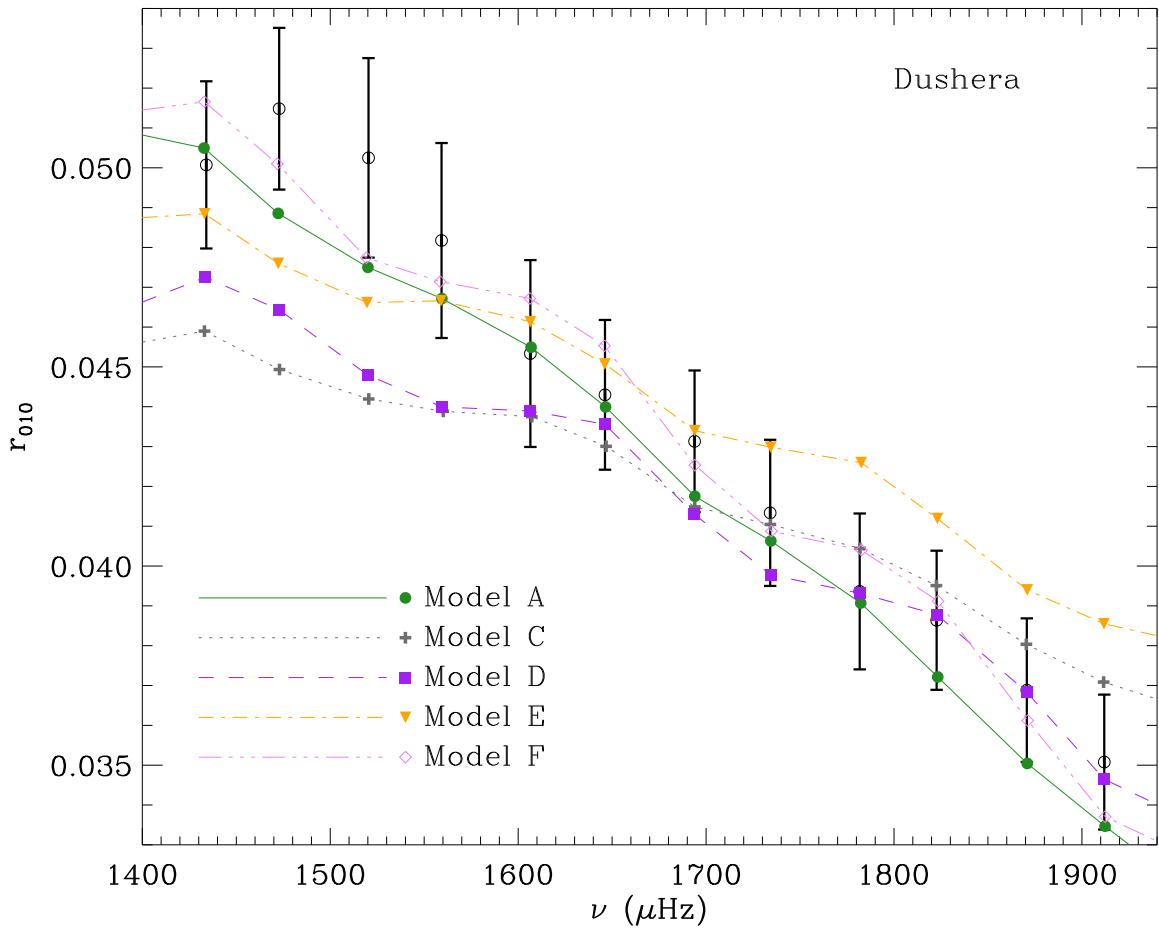}
\caption{Ratios $r_{010}$ of Dushera. Selected models plotted are A~(filled circles), C~(filled crosses), D~(filled squares), E~(filled downward triangles), and F~(open diamonds).}
\label{fig:rat_dush}
\end{figure*}

In Fig.~\ref{fig:rat_dush} we plot the ratios $r_{010}$ of the first 5 models in Table~\ref{tab:model_dush}. The slopes of models~C~and~E are clearly different from the observed one, with model~C~(E) showing lower (higher) values than the data in the low (high) frequency range of the considered ratios. Model~D presents a similar behavior as model~C.

Models~A and~F are the only two that consistently fit all the available data of Dushera, reproducing also the slope of the observed ratios $r_{010}$ and $r_{02}$. Considering only these results, our analysis points to a star with $R=1.39\pm0.01~R_\sun$, $M=1.15\pm~0.04M_\sun$, and $3.80\pm0.37$ Gyr corresponding to uncertainties of 1.0\%, 3.1\% and 9.9\% in radius, mass and age, respectively. As in the case of Perky we caution the reader that these values are to be taken as a reference, since systematics can arise from unexplored input physics and from models with slightly different properties that can also fit the data correctly but are overlooked by the fitting algorithms. We have shown that models fitting all asteroseismic combinations are restricted to a very small stellar parameter space and thus we expect the total uncertainties to be only slightly larger than those given above.

The analysis of Dushera not only reveals the presence of a convective core but it also puts constraints on its size. Our diagnostics are sensitive to the total extent of the mixed region, which in this case was reproduced by inclusion of overshooting. It is clear from Table~\ref{tab:ast_dush} and Fig.~\ref{fig:rat_dush} that only models with a mixed core comprising more than 2.35\% of the total acoustic radius are able to reproduce the ratios. The overall results leave us with a homogeneously mixed region of $\sim$2.4\%~$\tilde{t}/\tilde{t}_\mathrm{Phot}$, and in any case extending beyond the formal convective boundary due to the inclusion of overshooting. Thus, we measure here for the first time mixing beyond the formal convective boundaries in the core of a {\it Kepler} field main-sequence star.

Tests were made on the impact of the core size by running model~A with higher and lower overshooting efficiency values to produce different sizes of mixed zones. These models were not able properly to reproduce the ratios $r_{010}$ and show that, unlike the case of Perky, these ratios are mostly sensitive to the size of a convective core when this is already well developed during the main-sequence phase. We also performed tests using the model~F set of evolutionary calculations on the impact of reducing the considered number of ratios $r_{010}$ by two, and found no significant difference in the obtained solutions.
\begin{figure}[!ht]
\centering
\includegraphics[width=\linewidth]{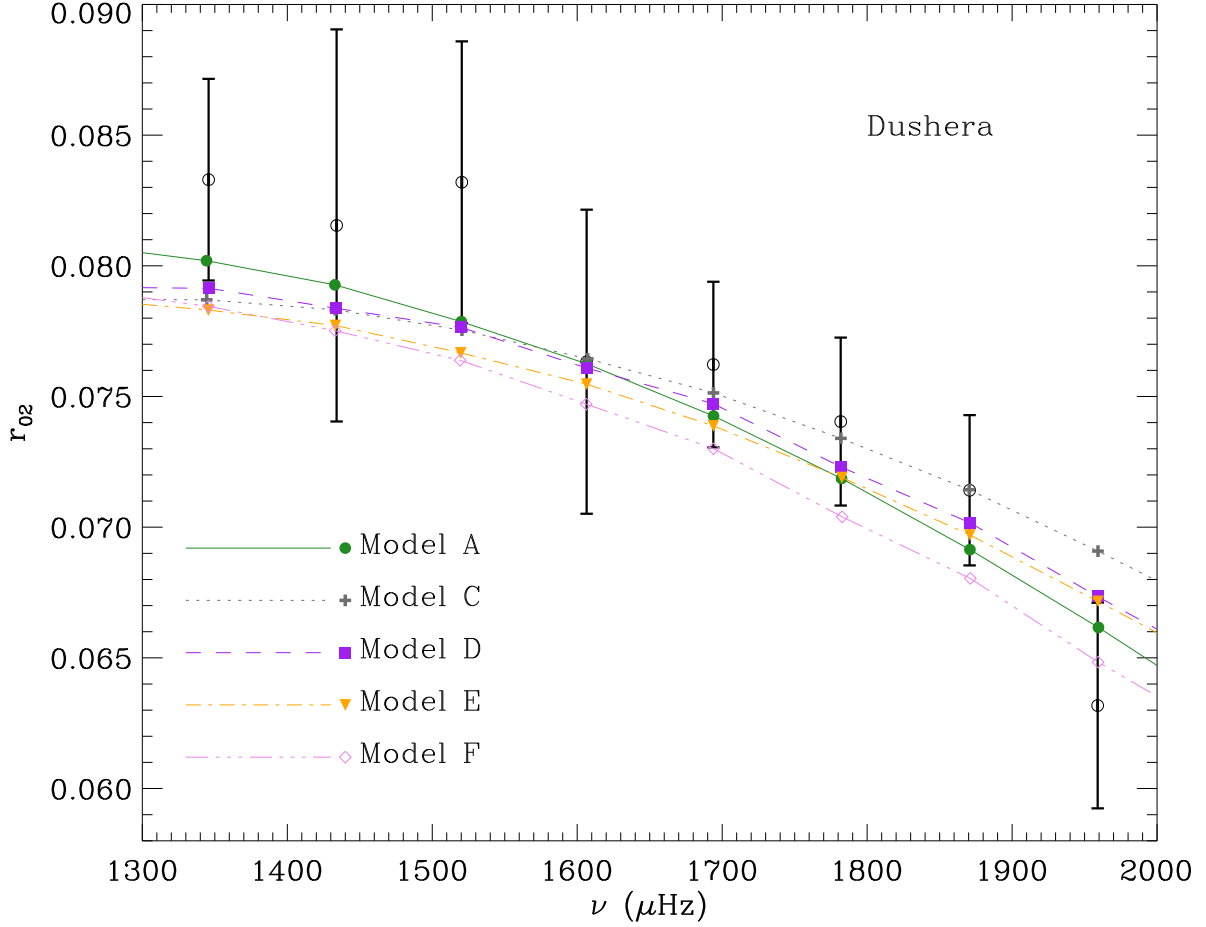}
\caption{Ratios $r_{02}$ for the data of Dushera. Selected models also plotted are A~(filled circles), C~(filled crosses), D~(filled squares), E~(filled downward triangles), and F~(open diamonds).}
\label{fig:rat02_dush}
\end{figure}
%
\section{Discussion}\label{s:disc}
Asteroseismic analysis aims to provide accurate stellar parameters of pulsating stars by theoretically reproducing their frequency spectrum. This has usually been done by matching the observed oscillation frequencies to theoretically calculated ones after applying an empirical correction to account for our poor modeling of stellar outer layers. Our results confirm that best-fit models found by this method do not always reproduce frequency combinations, and that different criteria to define what the best-fit model is can lead to solutions with similar stellar parameters but very different interior structures. As originally stated by \citet{Kjeldsen:2008kw}, the surface correction only ensures an accurate reproduction of the stellar mean density by assuming that the frequency offsets are caused solely by the properties of the outer layers. Usage of non-radial modes and frequency combinations is necessary to match properly the deep interior and provide a reliable estimate of stellar age.

Of particular importance is the implementation of the surface correction itself, which has a direct impact in the goodness of fit via the reduced $\chi^2_\nu$ value. The values of the corrections are much larger than the errors in the frequencies, making the definition of the best-fit model sensitive to the way these corrections are applied. Thus, the $\chi^2_\nu$ criterion cannot be objectively used among best-fits found using different implementations of the surface correction. We have proposed instead to characterize the goodness of fit to asteroseismic data by using the frequency ratios $r_{010}$ and $r_{02}$. These combinations are thought to be mostly determined by the central conditions of the star and almost unaffected by the outer layers, providing a more robust approach than matching individual frequencies. The masses and radii obtained by both techniques agree within 2-$\sigma$, but the resulting models can have very different structural properties. Thus, although the overall results are not hugely affected by the fitting techniques and the issue of the surface correction, using frequency ratios instead of just reproducing the individual frequencies of oscillation is a more reliable way to yield accurate stellar parameters, in particular age \citep[as was also suggested by][]{Miglio:2005is}.

The frequency ratios have been used as proxies for different internal properties of stars. For instance, a tight correlation between $r_{02}$ and central hydrogen content (age) holds for a given set of input physics, as shown by the modified C-D diagram in Fig.~\ref{fig:data_r02}. However, no correlation is found between these parameters among the set of models that fit the seismic data. The reason for this is that the equivalence between central hydrogen content and $r_{02}$ is only valid for a fixed set of input physics, which is naturally not the case in this study \citep{2002ESASP.485..291M,White:2011fw}.

In a similar way the ratios $r_{010}$, which were thought to be sensitive to the existence and size of the convective core as well as a proxy of the central hydrogen abundance, were found to be affected by other parameters not yet fully identified. The latter sensitivity is particularly important in the case of Perky, where analysis of $r_{010}$ did not allow us to discriminate between models with or without a convective core. The stellar parameters obtained at this stage are not significantly affected by the presence of a convective core; thus we can still derive a reliable mass, radius and age for this star. Nevertheless the remaining central hydrogen content of the target is not well constrained, and therefore neither is the age it will have at the end of the main sequence.

More importantly, the experience of Perky is a warning flag to our fitting algorithms and the use of the surface correction. It is easy to imagine that if all teams used a different implementation of the surface correction, different solutions from the ones presented here would have been found. In fact, the restriction to the amount of overshooting in the case of small convective cores might be overestimated in cases like Perky. It has been shown for the solar case that restricting the amount of pre-main sequence overshooting is necessary to reproduce helioseismic data \citep{1999A&A...347..272S}, but for stars slightly more massive than the Sun this needs to be further explored.

Related to fitting techniques, it remains fully to understand why in some cases low initial helium abundances are obtained when only individual frequencies are matched. Although the results for Perky suggest that mass is the major contributor to this feature, the overall picture is certainly more complex. While there is no reason to believe a universal value of $\Delta Y/\Delta Z$ applies for all stars, and small variations in helium are therefore expected, sub-SBBN values are clearly a challenge to stellar models. Such a behavior already appeared from the analysis of the main-sequence broadening in field K dwarfs \citep{Casagrande:2007ck}, and simple analytical considerations suggested that model opacities could have been a likely culprit \citep{2010MNRAS.406.1570P}.

\citet{Mathur:2012bj} performed a uniform asteroseismic analysis of 22 {\it Kepler} targets using AMP, the same code and fitting algorithm of models~E described in Section~\ref{sss:modE}. Their final results show that 10 of their 22 best-fit models have a mean value of initial helium below that of SBBN (see their Table~5). The models in \citet{Mathur:2012bj} are not directly comparable to ours since they are based on analysis of only 1-month long observations, but the overall outcome clearly points out to the need of further studies in this topic. These are currently underway (J\o rgensen et al. 2013, in preparation). Understanding these discrepancies is important not only for the sake of stellar models, but for the possibility of using these stars as tracers of the Galactic $\Delta Y/\Delta Z$ and possibly one day for multiple populations in stellar clusters

Stars like Perky lie at the limiting mass where the onset of core convection is thought to occur. Although such targets have been used in open clusters as a test for the solar abundances \citep{Vandenberg:2007cq}, it has been shown that the existence or absence of a convective core in stellar models heavily depends on our assumptions about the input physics. The upcoming releases of {\it Kepler} data will give us more frequencies with higher accuracy and we can certainly hope to solve this issue.

In the case of Dushera, we have made the first direct detection of a convective core in a {\it Kepler} main-sequence target. Use of fits to the ratios $r_{010}$ and $r_{02}$ allow us to discard outliers and determine the characteristics of the star with an excellent accuracy. The size of the central mixed region is estimated to be $\sim$2.4\% of the photospheric acoustic radius, showing that mixing beyond the formal Schwarzschild convective boundary exists. In this study the extra mixing has been modeled using different overshooting prescriptions; in reality it could also be produced, for instance, by rotational mixing or a combination of more physical processes.

For both targets, masses and radii obtained by our asteroseismic analysis have a precision of $\sim$4\% and $\sim$1.5\%, respectively. Regarding stellar ages, we determine them with an encouraging $\sim$10\% level of precision. This is probably optimistic due to our fitting algorithms potentially overlooking models that might better reproduce the data, and also since our models do not explore all possible sets of input physics. Thus, additional sources of uncertainties can exist that have not been explored yet. In fact, systematics arising from, for instance, the unknown input physics can easily account for an extra $\sim$7\% in age at the end of the main-sequence phase \citep{Valle:2012bz}. Although our analysis confirms that we are certainly doing better than the $\sim$20-50\% in age obtained with the usual isochrone fitting from a single stellar evolution database \citep[e.g.,][]{2007ApJS..168..297T,Soderblom:2010kr}, asteroseismology has the potential to deliver even better ages once we fully understand our current diagnostic tools and pursue further theoretical studies to develop even better ones.
\acknowledgements
The authors would like to thank the anonymous referee for useful suggestions and comments that improved the quality of the paper, and also Martin Asplund and Nicolas Grevesse for clarifying discussions about spectroscopic determination of stellar parameters. Funding for this Discovery mission is provided by NASA's Science Mission Directorate. The authors wish to thank the entire \emph{Kepler} team, without whom these results would not be possible. We also thank all funding councils and agencies that have supported the activities of KASC Working Group\,1. We are also grateful for support from the International Space Science Institute (ISSI). The authors acknowledge the KITP staff of UCSB for their warm hospitality during the research program ``Asteroseismology in the Space Age''. This KITP program was supported in part by the National Science Foundation of the United States under Grant No. NSF PHY05Ð51164. Funding for the Stellar Astrophysics Centre is provided by The Danish National Research Foundation (Grant agreement No. DNRF106). The research is supported by the ASTERISK project (ASTERoseismic Investigations with SONG and {\it Kepler}) funded by the European Research Council (Grant agreement No. 267864). V.S.A. received financial support from the {\sl Excellence cluster ``Origin and Structure of the Universe''} (Garching). S.B. acknowledges NSF grant AST-1105930. The National Center for Atmospheric Research (NCAR) is sponsored by the National Science Foundation. I.M.B., M.S.C., and S.G.S. acknowledge the support from the Fundac\~{a}o para a Ci\^encia e Tecnologia (Portugal) through the grants SFRH/BD/41213/2007, SFRH/BPD/47611/2008 and SFRH/BPD/84810/2012 and the ERC, under FP7/EC, through the project FP7-SPACE-2012-312844. T.S.M. acknowledges support from NASA grant NNX09AE59G. A.M.S is partially supported by the European Union International Reintegration Grant PIRG-GA-2009-247732, the MICINN grant AYA2011-24704, by the ESF EUROCORES Program EuroGENESIS (MICINN grant EUI2009-04170), by SGR grants of the Generalitat de Catalunya and by the EU-FEDER funds. W.J.C acknowledges support from the UK Science and Technology Facilities Council (STFC). This research was carried out while O.L.C. was a Henri Poincar\'e Fellow at the Observatoire de la C\^ote d'Azur, partially funded by the Conseil G\'en\'eral des Alpes-Maritimes. R.A.G. acknowledges the funding from the European CommunityÕs Seventh Framework Programme (FP7/2007-2013) under grant agreement No. 269194 (IRSES/ASK).
\bibliography{pe18}
\clearpage
\begin{table}[!ht]\scriptsize
\centering
\caption{Frequencies of Perky}\label{tab:frq_perk}
\begin{tabular}{c c}
\hline\hline
\noalign{\smallskip}
$\ell$ & $\nu\, (\mu$Hz) \\
\noalign{\smallskip}
\hline
\noalign{\smallskip}
0 &$ 1705.262\pm0.150 $\\
\noalign{\smallskip}
0 &$ 1807.695\pm0.195 $\\
\noalign{\smallskip}
0 &$ 1909.986\pm0.151 $\\
\noalign{\smallskip}
0 &$ 2012.862\pm0.132 $\\
\noalign{\smallskip}
0 &$ 2117.342\pm0.113 $\\
\noalign{\smallskip}
0 &$ 2221.532\pm0.103 $\\
\noalign{\smallskip}
0 &$ 2325.586\pm0.136 $\\
\noalign{\smallskip}
0 &$ 2429.842\pm0.151 $\\
\noalign{\smallskip}
0 &$ 2533.811\pm0.206 $\\
\noalign{\smallskip}
0 &$ 2639.023\pm0.351 $\\
\noalign{\smallskip}
0 &$ 2744.557\pm0.778 $\\
\noalign{\smallskip}
\hline
\noalign{\smallskip}
1 &$ 1752.393\pm0.191 $\\
\noalign{\smallskip}
1 &$ 1854.664\pm0.155 $\\
\noalign{\smallskip}
1 &$ 1957.109\pm0.125 $\\
\noalign{\smallskip}
1 &$ 2061.479\pm0.107 $\\
\noalign{\smallskip}
1 &$ 2165.840\pm0.110 $\\
\noalign{\smallskip}
1 &$ 2270.400\pm0.116 $\\
\noalign{\smallskip}
1 &$ 2374.555\pm0.139 $\\
\noalign{\smallskip}
1 &$ 2479.027\pm0.178 $\\
\noalign{\smallskip}
1 &$ 2583.464\pm0.213 $\\
\noalign{\smallskip}
1 &$ 2688.717\pm0.362 $\\
\noalign{\smallskip}
1 &$ 2794.513\pm0.515 $\\
\noalign{\smallskip}
\hline
\noalign{\smallskip}
2 &$ 1697.972\pm0.254 $\\
\noalign{\smallskip}
2 &$ 1800.063\pm0.392 $\\
\noalign{\smallskip}
2 &$ 1902.316\pm0.506 $\\
\noalign{\smallskip}
2 &$ 2005.719\pm0.184 $\\
\noalign{\smallskip}
2 &$ 2110.165\pm0.114 $\\
\noalign{\smallskip}
2 &$ 2214.715\pm0.202 $\\
\noalign{\smallskip}
2 &$ 2318.862\pm0.180 $\\
\noalign{\smallskip}
2 &$ 2422.900\pm0.288 $\\
\noalign{\smallskip}
2 &$ 2528.083\pm0.373 $\\
\noalign{\smallskip}
2 &$ 2631.927\pm0.700 $\\
\noalign{\smallskip}
2 &$ 2740.448\pm0.890 $\\
\noalign{\smallskip}
\hline
\end{tabular}
\end{table}
\end{document}